# A Perspective on High-performance CNT fibres for Structural Composites

*Anastasiia Mikhalchan\*, Juan José Vilatela\**

IMDEA Materials Institute, C/Eric Kandel 2, Getafe, Madrid, Spain, 28906

This review summarizes progress on structural composites with carbon nanotube (CNT) fibres. It starts by analyzing their development towards a macroscopic ensemble of elongated and aligned crystalline domains alongside the evolution of the structure of traditional high-performance fibres. Literature on tensile properties suggests that there are two emerging grades: highly aligned fibres spun from liquid crystalline solutions, with high modulus (160 GPa/SG) and strength (1.6 GPa/SG), and spun from aerogels of ultra-long nanotubes, combining high strength and fracture energy (up to 100J/g). The fabrication of large unidirectional fabrics with similar properties as the fibres is presently a challenge, which CNT alignment remaining a key factor. A promising approach is to produce fabrics directly from aerogel filaments without having to densify and handle individual CNT fibres. Structural composites of CNT fibres have reached longitudinal properties of about 1 GPa strength and 140 GPa modulus, however, on relatively small samples. In general, there is need to demonstrate fabrication of large CNT fibre laminate composites using standard fabrication routes and to study longitudinal and transverse mechanical properties in tension and compression. Complementary areas of development are interlaminar reinforcement

\* Corresponding Author. E-mail: anastasiia.mikhalchan@imdea.org (Anastasiia Mikhalchan)
\* Corresponding Author. E-mail: juanjose.vilatela@imdea.org (Juan José Vilatela)

with CNT fabric interleaves, and multifunctional structural composites with energy storage or harvesting functions.

TABLE OF CONTENTS

**PART 1. State-of-the-art of CNT fibres**

*1.1 Synthesis and hierarchical structure of CNT fibres*

Fibres [1-8], yarns [9-12], mats [13-17], and unidirectional textiles [18-19] formed of networks of thousands of carbon nanotubes (CNT) constitute a relatively new class of materials. Since they emerged on the arena of high-performance materials almost two decades ago [20], there has been continuous progress in the controlled synthesis of CNT fibres[1] at laboratory-scale quantities and beyond, and in the improvement of their properties. The application of CNT fibres spans from structural reinforcement fibres to porous electrodes and transparent conductors in optoelectronic devices. This review focuses exclusively on the first aspect.

Perhaps the first unusual feature of CNT fibres is the possibility to produce them by three methods very different from each other: wet-spinning from lyotropic liquid-crystalline phases of carbon nanotubes or polymer dispersions [1,2], dry-spinning of CNT fibres from CVD-grown vertical CNT arrays (so-called "forests") [4,5,9-12], and single-step direct spinning of CNT fibres from gas-phase Floating Catalyst CVD (FCCVD) [6-8].

CNT fibre wet-spinning can be achieved from spinnable dopes of highly-crystalline nanotubes dissolved in super acids such as fuming sulfuric or chlorosulfuric acid. The method is similar to that used to spin fibres of rigid-rod polymers such as polyparaphenylene terephthalate (PPTA) or polyparaphenylenebenzobisoxazole (PBO), traded as the well-known high-performance fibres Kevlar or Zylon [21]. CNT fibres produced by this method feature a very high degree of alignment and purity, and relatively dense packing of nanotubes with little porosity (density of 1.58 g/cm$^3$) [22]. The method offers, in principle, good control over CNT type, diameter, and aspect ratio because the CNT fibres can be spun from commercially available nanotubes, provided they are of high purity and high crystallinity [23]. The requirements of high purity can, however, make the starting CNT material currently expensive and have thus limited availability of large volumes required for composite studies. Dispersions of nanotubes in more conventional solvents or in polymers solutions have also been wet-spun into fibres, but the resulting fibres have typically lower degree of alignment and thus reduced properties, although offering higher potential for scalability.

---

[1] Nomenclature used in this paper: *CNT filament*: general description of a macroscopic continuous ensemble of nanotubes with preferential alignment parallel to the spinning direction. *CNT fibre* (also referred to as CNT yarn): individual, continuous, macroscopic filament, typically of 5-50 μm in diameter spun using one of the three main methods, and typically characterized after solvent densification and/or twisting. *CNT fabric*: unidirectional non-woven fabric produced by association of multiple CNT filaments, typically by filament winding before densification. *CNT fabric interleaf*: thin (< 50 μm) CNT fabric used as interlaminar reinforcement layer between two plies of a traditional laminate composites; also referred to as CNT veil in the literature.

Dry-spinning from vertical CNT arrays consists in pre-synthesizing of highly aligned CNT arrays grown on a flat substrate and then drawing the nanotubes from the side wall of such array, taking advantage of the van der Waals interaction between nanotubes. The spun material corresponds to a CNT network that is typically densified by applying twist, by capillary wetting upon exposure to a solvent, or by depositing as a thin layer on a substrate. Extensive optimization of the synthesis step has led to substantial control over the CNT diameter and length, combined with an expected narrow CNT length distribution and a low impurity content in the resulting fibre. This method has been widely used to produce CNT fibre composites and provided valuable insights into the structural factors governing composite mechanical properties. Despite extensive efforts to the scale up the "forest-spinning method", there remain concerns about fundamental challenges to produce large CNT fibre volumes continuously using this technique.

In the FCCVD method, also referred to as the direct spinning method, a fibre is produced by drawing a continuous CNT aerogel directly from the gas phase during growth of the nanotubes by FCCVD. The method takes the major advantage of combining the continuous synthesis and collection of the CNT fibres; however, this makes the control over these processes to be more complicated from the engineering point of view. Another drawback is the impurity content, which can be relatively high compared to other spinning methods, and, therefore, the rigorous optimization of the synthesis parameters and/or additional purification is necessary. Although the recent production at the laboratory-scale remains low, the method is considered inherently suited for mass-production, offering the continuous and fast spinning of nanotubes of very high aspect ratio [7,8]. Yet, the research on the practical control over CNT length and type (semiconducting or metallic) remains on-going [24].

The CNT fibre spinning methods discussed above are similar to wet-spinning of rigid-rod polymer fibres, dry spinning of staple fibre, and synthesis of vapour grown carbon fibres, respectively. That the same CNT fibre can show features of such different classes of materials is very unusual and interesting. Central to this multifaceted nature of CNT fibres is their hierarchical structure. Figure 1 shows the structure of CNT fibres produced by FCCVD at different magnifications. The basic building blocks are individual nanotubes, which associate in bundles of a few nanometres in diameter. The CNT aerogel drawn continuously out of the reactor can be densified as a single fibre (Figure 1c) or as a unidirectional non-woven CNT fabric (Figure 1d). As discussed further below, the resemblance of CNT fibres to other high-performance fibres is through features present at different length-scales. Broadly speaking, their transport properties are reminiscent of the graphitic structure of the building blocks, their similarity to polymer fibres is related to their bundle network structure, and their yarn-like character most clearly observed from the filament level upwards.

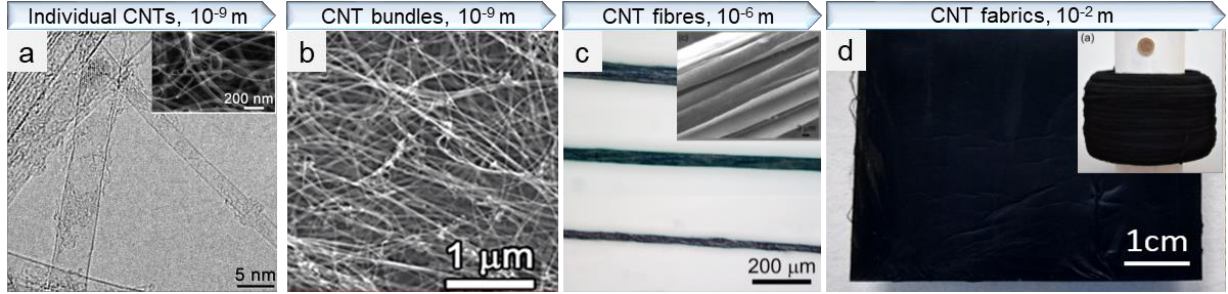

**Figure 1.** Hierarchical structure of CNT fibres and fabrics at different length scales: a) TEM image of collapsed individual nanotubes, inset: SEM image of nanotubes accumulated in bundles (adapted with permission from [25]. Copyright 2014 American Chemical Society); b) SEM image of an assembly of thousands of CNT bundles (adapted with permission from [8]. Copyright 2015 American Chemical Society); c) optical image of continuous CNT fibres produced by FCCVD and densified with acetone, inset: SEM image of an individual CNT fibre with a non-circular cross section (adapted with permission from [25]. Copyright 2014 American Chemical Society); d) photograph of a macro-scale CNT fabric made of thousands of aligned CNT fibres (image courtesy by J. J. Vilatela), inset: a bobbin with about 1 km of non-densified CNT fibre produced over a couple of hours of continuous spinning (adapted with permission from [25]. Copyright 2014 American Chemical Society).

*1.2 Mechanical properties of CNT fibres*

Figure 2 summarizes the recent examples of tensile strength and Young's modulus of CNT fibres synthesized via different methods and without post-spinning treatments. Note that these properties are represented in specific units, either in N/tex or GPa/SG (GPa/Specific Gravity). The cross-section of CNT fibres is typically not circular, giving a large source of errors in ultimate strength or modulus when in GPa units [26]. A comparison of the tensile properties in N/tex rather than in GPa is also beneficial due to the intrinsically low-density of CNT materials, typically below 1 g/cm$^3$, which is lower than that of classic high-performance fibres (density of commercial aramid fibres is 1.44 g/cm$^3$; PBO 1.56 g/cm$^3$, carbon fibres 1.7 – 2.1 g/cm$^3$, and S-glass fibres 2.58 g/cm$^3$, respectively).

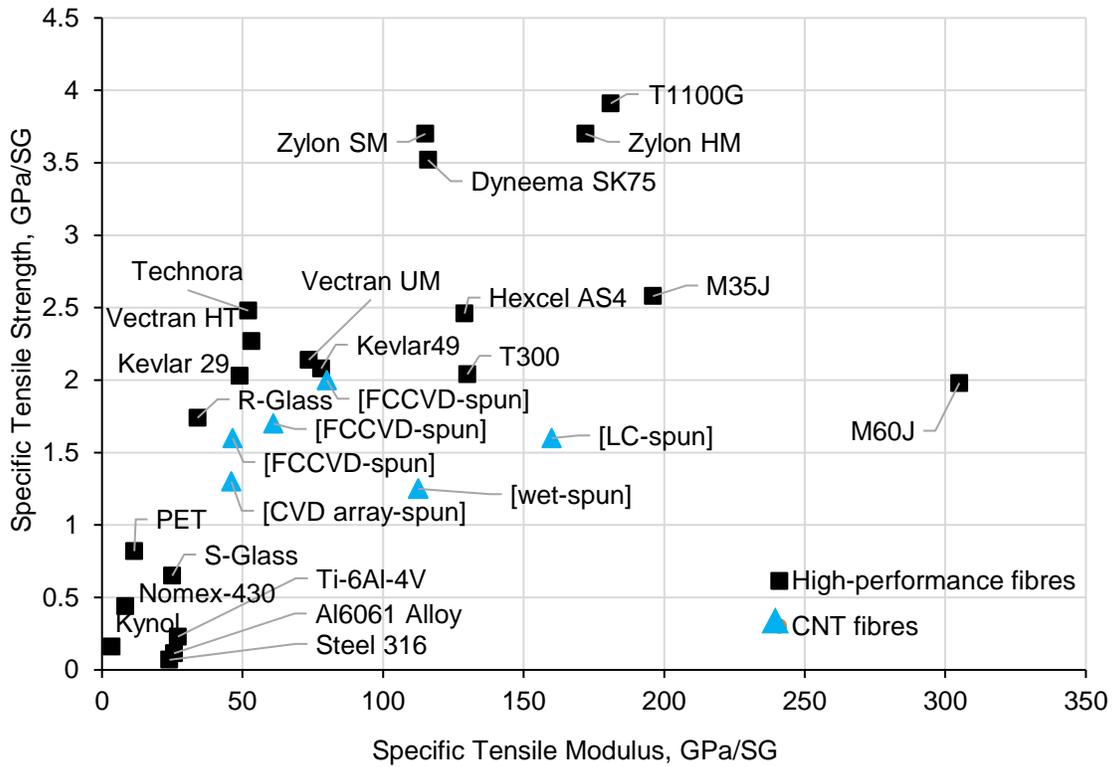

**Figure 2.** Specific tensile strength *versus* specific Young's modulus of CNT fibres compared with commercial high-performance materials. Datapoints for CNT fibres are taken from [7,23,27-30].

Nowadays, the specific tensile strength and elastic modulus of as-synthesized CNT fibres are roughly at around 1.5 N/tex (for the FCCVD spinning) and 70 GPa/SG, respectively. The highest elastic modulus of 160±40 GPa/SG has been demonstrated for CNT fibres synthesized from liquid crystalline (LC) solutions of commercially available carbon nanotubes of high quality and 4,400 aspect ratio [23]. In terms of specific values, CNT fibres already outperform widely-used steel and Al alloys, as well as PET, meta-aramid (Nomex), and glass fibres (GF). An increasing number of examples of CNT fibres are reaching 2 GPa/SG strength and 100-150 GPa/SG stiffness, almost at the level of liquid crystalline aromatic polyester (Vectran), para-aramid (Kevlar), and medium-strength commercial carbon fibres (T300).

A relevant, seemingly inherent feature of CNT fibres is their high fracture energy in tension (Figure 3). It embodies the general high toughness of the material, which facilitates handling and further processing. But importantly, the high density-normalized fracture energy values in the range of 50-100 J/g, are also desirable for the applications where energy absorption and damage tolerance are of the essence, leading for example to a high figure of merit for ballistic protection [27].

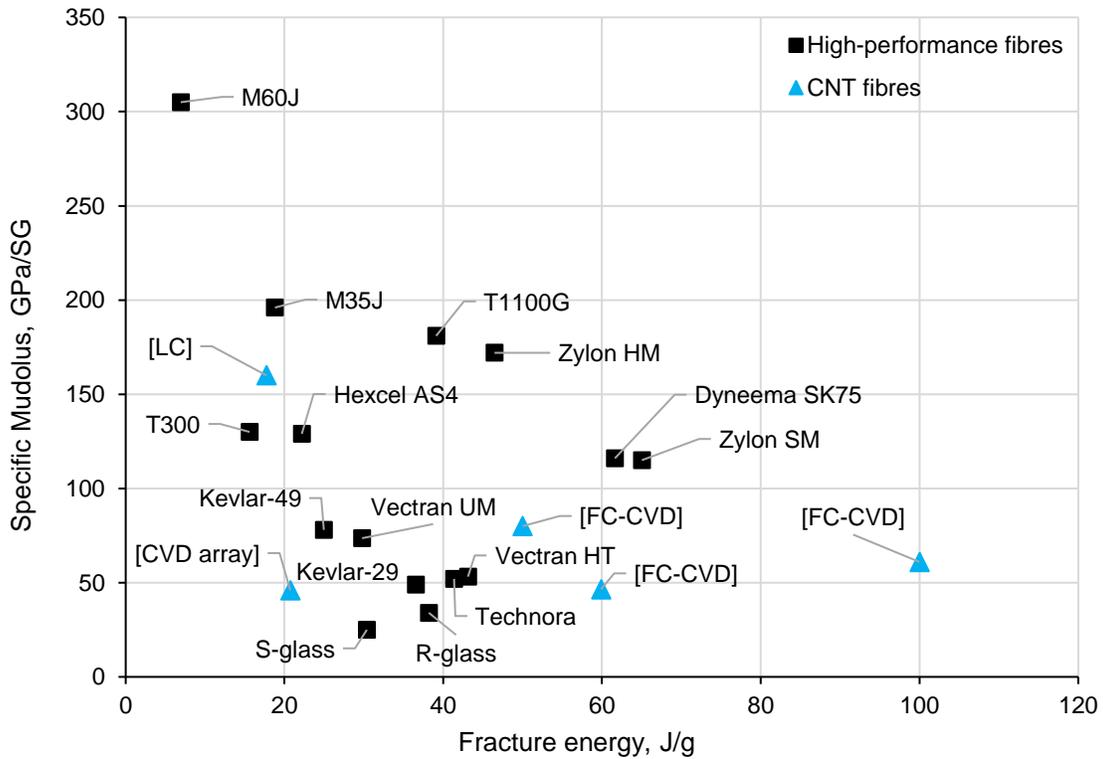

**Figure 3.** Specific tensile modulus *versus* fracture energy of CNT fibres compared with commercial high-performance fibres. Datapoints for CNT fibres are taken from [7,23,27-29].

However, mechanical properties of as-synthesized CNT fibres (note, without post-treatments) are still far below that of commercial ultra-high-molecular-weight polyethylene (UHMWPE) fibres (Dyneema, Spectra), high-modulus and high-strength carbon fibres (T1100GC, M60JB), and fibres made of rigid-rod polymers (Zylon). More importantly, current tensile properties of CNT fibres are still a small fraction of the tensile strength and modulus of individual CNT bundles, recently measured at 27-31 GPa/SG and 337-640 GPa/SG for bundles comprised of 2-7 nanotubes [31]. This indicates that there is large room for improvement. In order to get an idea about the extent and the time frame of potential improvements in tensile properties the next section briefly reviews the historical development of CNT fibres alongside that of carbon fibres (CF) and UHMWPE fibres.

*1.3 Historical retrospective on R&D of commercial high-performance fibres and CNT fibres*

The first work on carbonized cellulosic filaments dates back to Thomas A. Edison's incandescent lighting experiments and W. R. Whitney's report of a "highly successful" process for graphitizing cellulosic filaments at temperatures higher than 2300 °C, more than a century ago [32]. The modern era of commercial carbon fibres, however, began in the middle of 1950s with efforts to develop a novel carbonaceous material which possesses unique physical properties and retains the characteristic textile attributes of the cellulosic precursor (e.g., fibre, yarn, cloth). Early samples were filamentary graphite [33,34], probably similar to VGCF or large-diameter nanotubes, with reported tensile strength approaching 20 GPa [35].

As captured in the first patents of 1958-1964, efforts were initially focused on developing a process to manufacture a continuous macroscopic fibre processes rather than on properties [36-50]. The tensile strength of the first-generation rayon-based carbon fibres in 1959 was below 77 MPa [38]. Even after improvements in techniques to measure tensile properties of individual 5-25 μm diameter "fibrous graphite" their tensile strength of 276 MPa was comparable to commercial lamp monofilaments of 146 μm in diameter with strength of 228 MPa [39].

A key milestone in the development of CF came from Tang and Bacon's studies on the structural rearrangements in cellulose fibres at high temperature treatments and the relation between the crystal structure of cellulose and graphite [51,52]. Their observation that the cellulosic crystalline structure started to break down at 305°C, prompted them to apply tensile stress parallel to the fibre axis during carbonization to favour alignment of graphite basal planes. Stretch-graphitisation increased tensile strength and modulus to 1.21 GPa and 150 GPa, respectively (Figure 4) [53].

Soon after, it was recognized that direct stress-graphitization of rayon-based fibre was limited by the strength of a carbonized fibre, making the processes ineffective and "not conductive to a successful commercial operation" [21]. Then followed extensive work on studying combinations of heat-tensioning treatments at the oxidation, stabilization, carbonization and graphitization stages. Strength and modulus of Thornel25 and Thornel50 rayon-based CF commercialized in the 1970s, for example, went from 1.24 GPa and 161 GPa to 1.93 GPa and 359 GPa, respectively, after stress-carbonization followed by stress-graphitization (Figure 4). Overall, the large number of CF grades reflects the complex relation between processing and properties, but an important lesson from CF manufacture is the need to combine stretching/tension with annealing methods [54]. The different tension-annealing steps depend strongly on the choice of polymer fibre precursor amongst rayon, PAN or mesophase pitch-based fibre, but a common objective of these methods is the development of a microstructure of extended, ordered domains predominantly aligned parallel to the fibre axis.

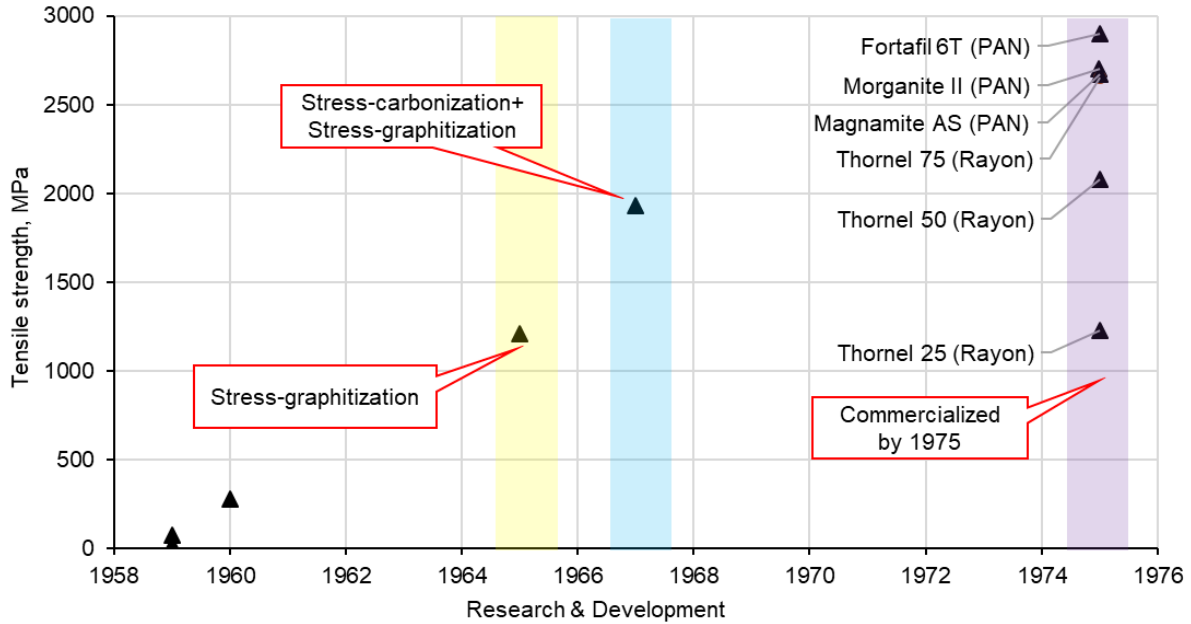

**Figure 4.** Historical retrospective on the first years of research and development of carbon fibres showing the significant improvement of mechanical properties followed the technological optimization.

There is also an interesting parallelism between the historical development of CF and UHMWPE fibres. After first reports on the impressive tensile properties of individual single-crystals of PE approaching tensile modulus of 240 GPa [55,56] and about 14 GPa strength [56], continuous fibres had unimpressive mechanical performance as a consequence of alternating crystalline and amorphous regions within the fibres. Pennings famously described this structure containing only a very small fraction of fully extended polymer macromolecules as 'shish-kebabs' [57-60]. The importance of uniaxial extensional flow to produce molecular alignment and chain extension and thus improved mechanical properties was clearly established [61], but in the 1970s PE fibre samples spun from melt and solid-state extrusion still exhibited strength of about 500-1500 MPa and modulus of 60-70 GPA [62,63]. The breakthrough happened with research on gel spinning and X-ray analysis of corresponding molecular arrangements by Smith and Lemstra [64] and Kalb and Pennings (Figure 5) [65]. Their work provided experimental evidence that gel PE fibres could be spun from dilute solutions of high molecular weight polymer with reduced number of entanglements per polymer chain, which could then be hot-drawn to produce fibres with high alignment and chain extension. Hot-drawing to ratios > 30 enhanced the tensile strength and stiffness from 10 MPa and 90 MPa for an initial gel-spun fibre, to 3 GPa and 90 GPa [64] and 3.7 GPa and 120 GPa [65], correspondingly. This method forms the basis for the manufacturing process of Spectra 100 PE and Dyneema SK60 fibres, commercialized since 1985-1987.

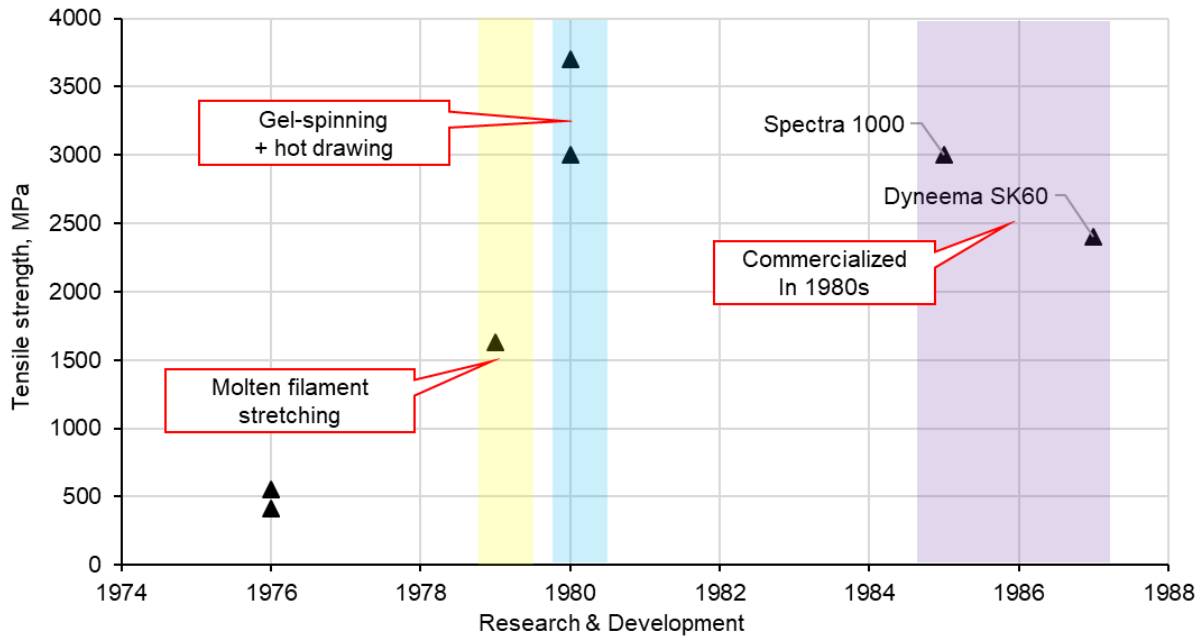

**Figure 5.** Historical evolution of the tensile properties of UHMWPE fibres showing milestone improvements in fibre spinning and structure, followed by their technological implementation.

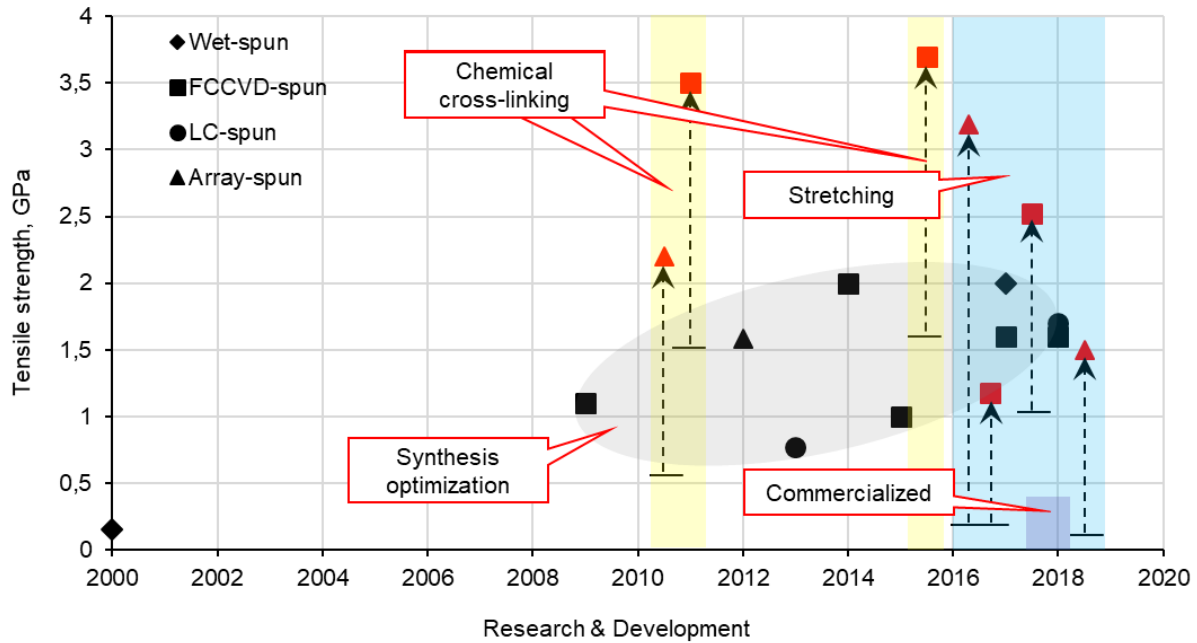

**Figure 6.** Historical evolution of the tensile properties of CNTF showing large property improvements arising from stretching and functionalization. Included some emerging data for commercial CNT fabrics.

A similar analysis of the evolution of tensile strength of CNT fibres (CNTF) is presented in Figure 6. After the seminal paper by Poulin's group [20] reporting tensile strength and stiffness of about 150 MPa and 9-15 GPa, respectively, there was rapid emergence of other CNT fibre spinning processes and improvement of tensile properties. Early work in the field was characterized by a large scattering in reported tensile properties. As CNT fibre spinning processes were better controlled and sample uniformity/reproducibility improved, tensile properties have narrowed down. This is particularly relevant for the fabrication of composite specimens, which require a large amount of CNT fibre material and whose properties can only be accurately analysed if the reinforcement material is reasonably uniform. Thus, in Figure 6 we include data for CNT fibres at large gauge lengths, and which show indication of continuous spinning of sufficiently large amounts for composite studies.

CNT growth by FCCVD occurs very fast, and typical collection rates at the laboratory setups could reach 1-2 km/h, giving a few kilometres of fibres per day of spinning but which corresponds to a few grams [7,8]. In wet-spinning from liquid crystalline solutions in acids, CNT fibres could be spun concurrently by using a multihole spinneret. The continuous spinning of 19 parallel CNT fibres from superacid solutions at the rate of 600 m/h has been demonstrated in 2013 [1]. There have been similarly successful efforts to increase production rates and develop water-based wet-spinning methods [66]. As a side note, it is interesting that in contrast to traditional development of synthetic fibres, CNT fibre property development efforts are concentrated in non-industrial research laboratories. Over the last decade, a few companies have undertaken the commercialization of CNT fibres and fabrics. The limited information available suggests that they have focused on scale up and the production of unidirectional non-woven fabrics, rather than fibres, although FCCVD fibres of 0.189 GPa strength and 9 GPa stiffness [67] have been also reported with claims that they can reach 4 GPa and 90 GPa, respectively, through a series of post-treatments (Nanocomp) [68].

Unidirectional CNT fabrics, also referred to as sheets or mats, are produced by overlaying multiple CNT aerogel filaments, which is a convenient way to produce free-standing macrcoscopic structures. However, literature values for mechanical properties of these fabrics are much lower than those of laboratory fibres: 30-45 MPa tensile strength and about 500 MPa stiffness (Tortech) [69]; 120-130 MPa strength and 1.5-1.7 GPa stiffness (Nanocomp) [70,71]; around 300 MPa strength and 5 GPa stiffness (Suzhou) [72].

The data in Figure 6 give testimony of a general improvement in tensile properties of CNT fibres since they were first reported, with current values firmly established in the high-performance range. Increases in modulus and tensile strength have been mainly a consequence of developments in synthesis, stretching and post-spin treatments. In general, property improvement has progressed faster than the understanding of the causes behind it, particularly with respect to the relative contributions of molecular composition, alignment, and stress transfer to bulk properties. Below, we summarise progress in finding suitable descriptors to characterise CNT fibres and transition from trial-and-error recipes to robust processing-structure-property models.

**PART 2. Molecular structure and mechanical properties of CNT fibres**

An interesting aspect of CNT fibres is their multifaceted nature with respect to structure and properties. Part of the exceptionally wide range of properties of CNT fibres stems from their high charge mobility and thermal conductivity, which are unusual in high-performance fibres. Thus, envisaged applications of CNT fibres spread beyond the structural realm to light-weight conductors, porous/catalytic electrodes, and purification membranes. But even from a purely structural and mechanical perspective the classification of CNT fibres is elusive, as they combine features of rigid-rod polymers, ultra-high molecular weight PE, CF, and staple yarns.

*2.1 Molecular composition: CNT as the "ultimate polymer"*

High-performance polymers have a sufficient number of chemical and physical bonds in the macromolecular structure for effective stress-transfer along the fibre's axis, while their stiffness is governed by the degree to which the chemical bonds are aligned [73]. Seen alongside common synthetic polymers used in fibres, carbon nanotubes could be considered as highly conjugated molecules similar to rigid-rod polymers (Figure 7).

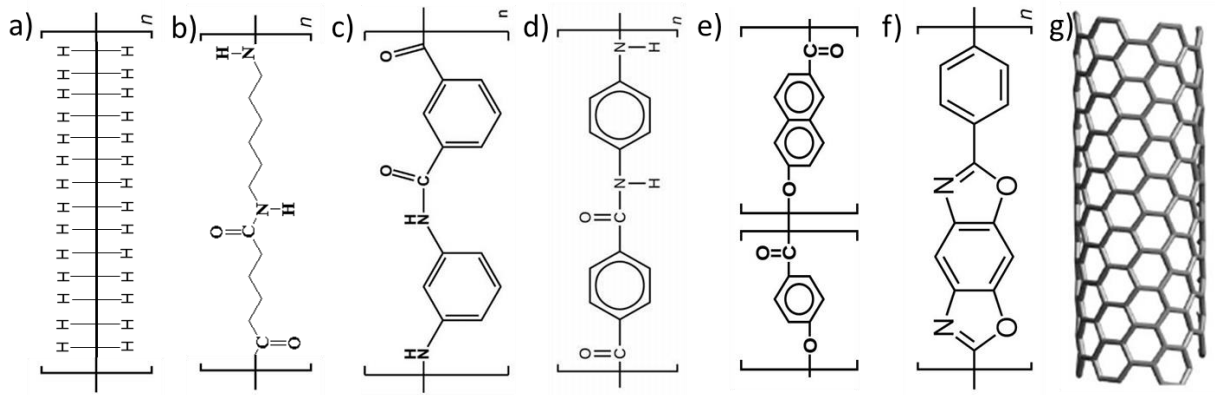

**Figure 7.** The examples of molecular structure: a) polyethylene; b) nylon 6; c) meta-aramid (Nomex); d) para-aramid (Kevlar); e) LC aromatic polyester (Vectran); f) PBO (Zylon), and g) a single-walled CNT as the "ultimate" polymer.

By replacing the aliphatic chain in nylon with benzene ring, the aromatic polyamide molecule appears with aromatic rings linked by amide groups in meta- or para-configuration (Figure 7 c,d). Such polymer molecules have higher elastic modulus than those of typical flexible chain polymers as PE, PP, nylon 6, and others. The aramid chains have fewer chain valence angles that can easily bend under mechanical load, thus, resulting in higher tensile performance of the fibres. Due to their high molecular stiffness, these polymers are processed from their liquid crystalline phases. The rigid-rod PBO molecule (Figure 7 f) is highly-conjugated, with the interatomic bonding being much stronger than the single covalent bonding in PE chain [73]. The PBO chain has no freedom for conformational mobility of any chemical bonds along the backbone (except the axial rotation), and the conjugated rings supply the overall rigid-rod shape of the molecule. This chemical structure is at the core of the tensile performance of the PBO fibres (commercialized as Zylon), with almost doubled the strength of para-aramids.

Many researchers have compared the molecular structure, properties, and processing of individual carbon nanotube with those of high-performance polymers, specifically, para-aramid and PBO [74-77]. Indeed, a nanotube could be considered as an example of the "ultimate polymer" (Figure 7 g), characterized by a molecular structure consisting of a repeating hexagonal mesh of $sp^2$-hybridized carbon atoms, rolled into a cylinder. Pasquali *et al.* have argued that single-walled nanotubes possess unusual properties that are "essentially extremes in the spectrum of polymer properties" [75]. As an example, the SWCNT persistence length determined by fluorescence visualization in aqueous suspensions is as high as 174 µm [78], which is extremely long compared with liquid crystal polymers (PPTA 15-30 nm [79], PBO 60-120 nm [75,80]). Similarly to stiff rod-like and rigid-rod polymers [81], nanotubes possess superior thermal stability, do not have the melting point or glass transition temperature, and are insoluble in most solvents, yet could be processed into fibres from the liquid crystalline solutions in super acids [1,2]. But it is noteworthy that whereas breaking of a polymer chains requires cleavage of a single covalent bond, the CNT molecular structure features covalent bonds in both the $x$ and $y$ directions, therefore, requiring cleavage of ten or more bonds [82,83]. Such conjugated structures makes nanotubes stable even in the presence of defects in the hexagonal plane [84,85]. This contrasts also with carbon fibre, which fails in a brittle manner and has low damage tolerance [86,87].

*2.2 Molecular interaction: fibrillary fracture*

In CNT fibres and similar macromolecular solids bulk properties are often dominated by inter, rather than intra molecular properties. The axial mechanical properties (particularly, strength) of highly-aligned polymeric fibres are reminiscent of the strong covalent bonds of the polymer chains, but because the chains are several orders of magnitude shorter than the macroscopic fibre, stress is inevitably transferred between adjacent chains by intermolecular bonds, either van der Waals or (more preferably) hydrogen bonds [73]. For example, neighboring para-aramid macromolecules interact through hydrogen bonds, whereas adjacent PBO molecules are linked only with van der Waals bonds, which are the source of their low shear strength and modulus and low compressive yielding strength. In contrast, the structural "building" elements in carbon fibres (extended graphite planes formed of hexagonally arranged carbon atoms) are coupled via covalent bonding, which is purely elastic and do not undergo plastic deformation under stress, making CF brittle in nature.

A weak molecular interaction in a fibre, coupled with long and thin constituent elements, leads to high flexibility in bending. This yarn-like character is often captured by tying an overhand knot in a single filament, and sometimes characterized by comparing strength with and without the knot. For a brittle material like CF failure occurs while tying the knot [88], at radii from tens of microns to 1-2 mm depending on grade. Most other high-performance fibres are tolerant to knotting due to the physical nature of interaction between the adjacent molecules (Figure 8 a-c).

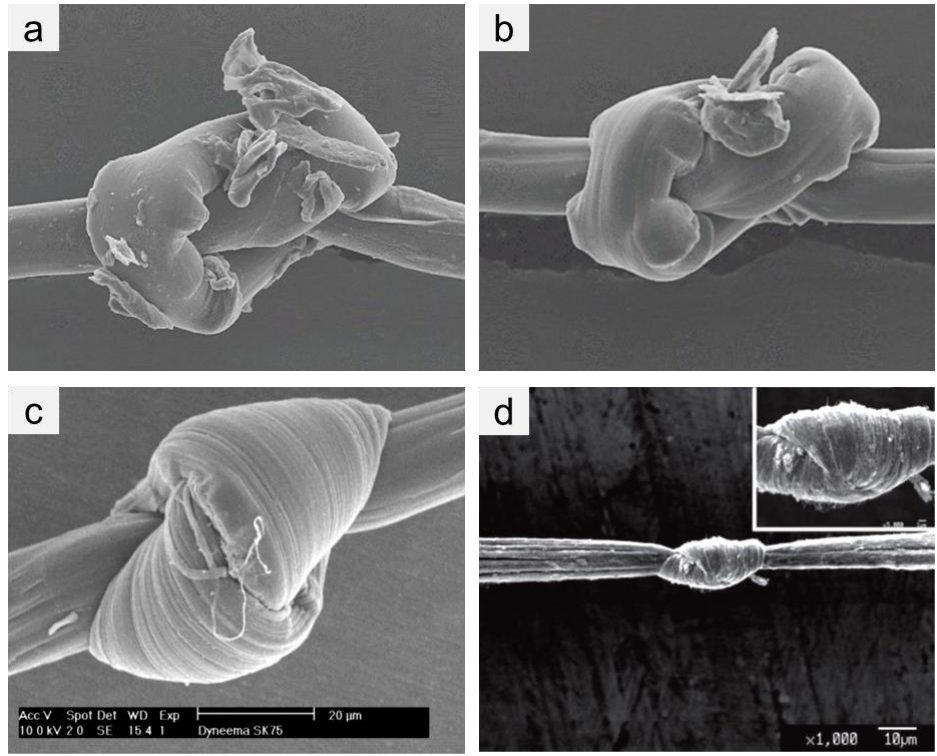

**Figure 8**. SEM images of the high-performance fibres with knots: a) LCP (Vectran) (https://eurofibers.com/ Copyright 2011 EuroFibers B.V.); b) PBO (Zylon) (https://eurofibers.com/ Copyright 2011 EuroFibers B.V.); c) UHMWPE (Dyneema) (adapted with permission from [89]. Copyright 2011 Scientific Research Publishing); d) CNT fibre (Inset: the knotted section of the CNT fibre is roughly 24 μm long but only 6 μm wide, showing the flattening of the fibre) (reproduced with permission from [90]. Copyright 2010 WILEY-VCH Verlag GmbH & Co.).

Due to the extended flexible chain nature and weak van der Waals interactions between its constituent molecules, UHMWPE fibre handles bending and knotting much easily compared to stiffer aramid and rigid-rod PBO fibres [91]. The latter experience multiple kinking and other structural damage under bending deformation, leading to substantial deterioration of mechanical properties. The tensile knot strength and stiffness efficiency (ratio of the strength of a knotted fibre to that of an unknotted control fibre expressed in %) for UHMWPE (Dyneema) was observed at the level of 50%, whereas for either para-aramid (Kevlar) or PBO it was only about 20% [90,92].

In spite of the compositional similarity with stiff and rigid-rod polymer fibres, the weak intermolecular forces between nanotubes in fibres makes them more similar to UHMWPE fibres. Tying a knot in a CNT fibres causes no apparent damage (Figure 8 d), and when the bending stress is released, the fibre can be straightened again with a very small axial stress and preserving almost 100% tensile strength. Similarly, when such CNTF is stretched across the sharp cutting edge of a razor blade (vertical plane), it spreads laterally (Figure 9 a), providing significant resistance to cutting and crack propagation through crack deflection [90]. Indeed, the weak intermolecular interactions between 30-40 nm microfibrils in UHMWPE fibre [91] makes it also spread laterally across a razor blade (Figure 9 b).

For both PE fibres [73] and CNTF [27,90,93,94] the shear strength between load-bearing elements is orders of magnitude smaller than the covalent bond strength of the macromolecular backbone, which implies that tensile failure occurs through extensive shearing of elements [95]. Examples of fibrillary fracture surfaces of PE fibres and CNTF spun by the FCCVD process are shown in Figure 9 (c, d).

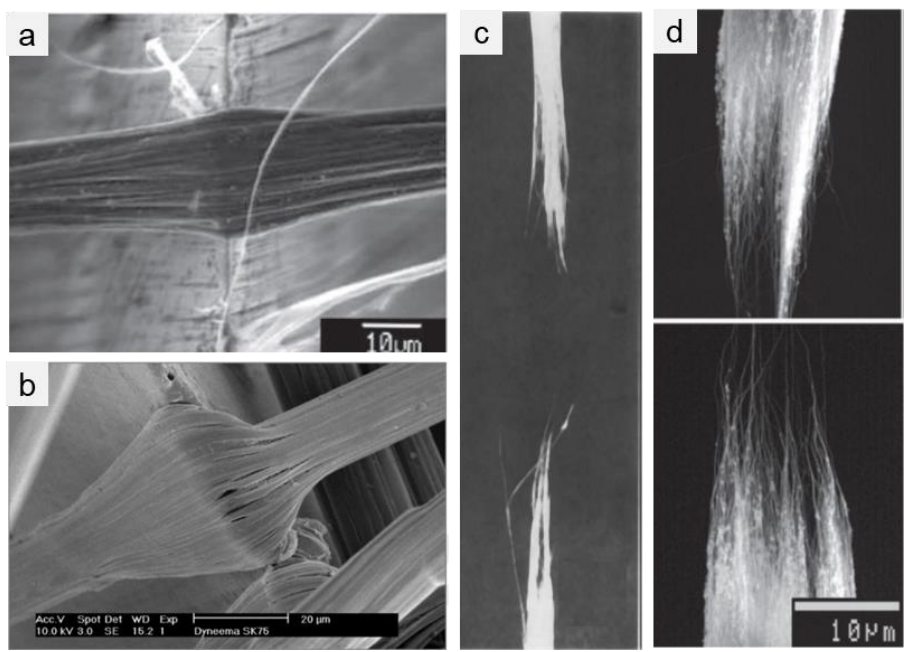

**Figure 9.** SEM images of the a) CNT fibre (adapted with permission from [90]. Copyright 2010 WILEY-VCH Verlag GmbH & Co.) and b) Dyneema fibre (adapted with permission from [89]. Copyright 2011 Scientific Research Publishing) pressed over the edge of a razor blade. Electron micrographs showing the distinctive fibrillar fracture mode for c) PE fibres (adapted with permission from [95]. Copyright 1972 WILEY-VCH Verlag GmbH & Co.) and d) CNTF (reproduced with permission from [90]. Copyright 2010 WILEY-VCH Verlag GmbH & Co.) through shear between the structural elements.

*2.3 CNT fibres as a fibrillary network of crystals*

The discussion above highlights the fact that CNT fibres are similar to CF with respect to the chemistry of their building blocks (and associated transport properties), to rigid-rod fibre in terms of the high stiffness of their constituent molecules, and to UHMWPE from the point of view of their weak intermolecular forces. But all these fibres share a common network structure of extended crystalline domains, highly-aligned parallel to the fibre axis. The basic structure of the "ideal continuous" polymer crystal fibre predicted in the 1930s by Staudinger [96-98], explains the origin of their high strength and stiffness (Figure 10 a), with subtle differences depending on the type of fibre. UHMWPE fibre is an assembly of microfibrils as crystalline blocks connected by tie molecules [91], similar to the crystal model proposed by Fisher (Figure 10 b). PAN-based carbon fibres can be described through a "ribbon-like" model in which graphite basal planes are packed side by side forming crystallites as a kind of microfibrils [99]. The simplified representation of the complex three-dimensional arrangement of undulating crystallites is shown in Figure 10 c. The structure of CNT-based fibres can be treated as a network of oriented crystalline domains, where these domains correspond to bundles of CNT closely packed at distances close to turbostratic graphite [27] (Figure 10 d).

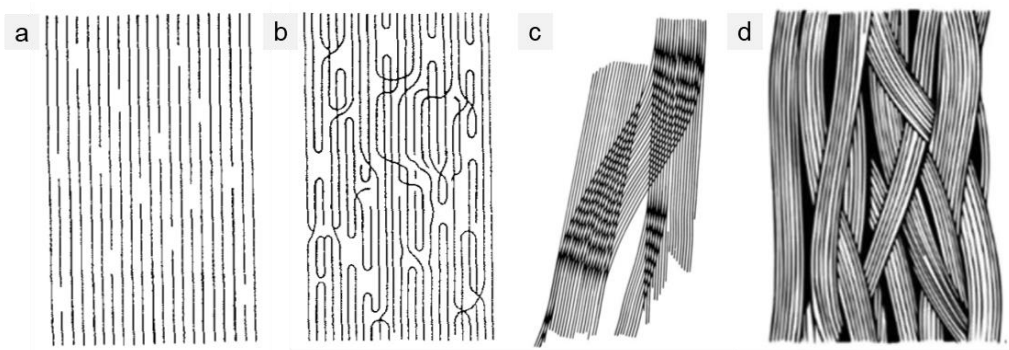

**Figure 10.** The fibrilar molecular model applied to different high-performance fibres: a) idealized continuous crystal for a polymer with perfect uniaxial extension of chains, as proposed by Staudinger (adapted with permission from [100]. Copyright 1995 Taylor & Francis); b) a polymer fibre comprising crystalline domains, structural defects and tie molecules, proposed by Fisher and Goddar (adapted with permission from [100]. Copyright 1995 Taylor & Francis); c) schematic two-dimensional representation of longitudinal structure of PAN-based carbon fibres with a misoriented crystallite linking two crystallites parallel to fibre axis (reproduced with permission from [101]. Copyright 1987 IOP Publishing Ltd.); d) schematic illustration of fibrillar morphology of a CNT fibre formed of aligned CNT bundles (adapted with permission from [27]. Copyright 2019 Elsevier Ltd.).

The Staudinger's model outlines the three constitutive requirements for high fibre strength/stiffness: 1) high molecular weight of long chains, with few chain end defects; 2) perfect orientation of chains along the axial direction; 3) perfect lateral packing of the chains [98].

An early hypothesis suggested that stress transfer between nanotubes in a fibre is similar to frictional stress transfer between primary filaments in staple fibre and thus, CNT fibre strength should scale with CNT length (Figure 11 a,b). More recently, Tsentalovich *et al.* carried out an extensive study of LC-spun fibre produced from a wide range of single-walled and double-walled nanotubes [23]. The samples are highly aligned and can be assumed to have the same degree of orientation. For both single-walled and double-walled nanotube fibre strength was found to scale with aspect ratio to the power of 0.9 [23], which is close to the linear dependence expected for stress transfer in shear for very long elements. A linear dependence of tensile properties and longitudinal electrical conductivity has also been observed for fibres spun from aqueous dispersions of nanotubes of different length and number of layers [102].

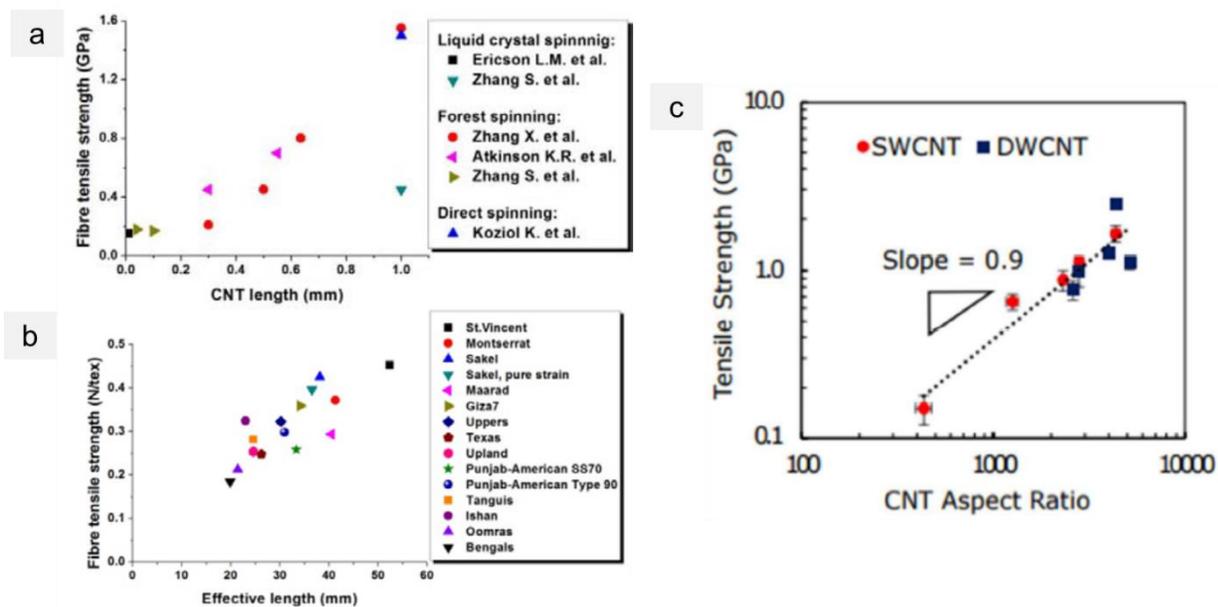

**Figure 11.** Dependence of fibre strength on length of the constituent nanotubes: a) early data including different processes (reproduced with permission from [76]. Copyright 2012 SAGE Publications) with b) similar behaviour to cotton varieties (reproduced with permission from [76]. Copyright 2012 SAGE Publications), c) nearly linear scaling of LC-spun fibres produced from single-walled and double-walled nanotubes of different lengths (adapted with permission from [23]. Copyright 2017 American Chemical Society).

These results highlight the interest in increasing length of CNT in fibres to achieve higher longitudinal tensile and transport properties. However, we note that in fact, the CNT length in LC-spun (up to 4.4 μm) and FCCVD-spun (1mm) fibres is already much higher than the typical length of aramid or PBO molecules of about 100 - 200 nm [75,103], or the length of crystalline blocks in Dyneema microfibrils of around 600-800 nm [91]. The need for unusually high aspect ratios to achieve high bulk tensile properties is partly a consequence of the low shear strength of graphitic systems; a motivation for post-spin treatments (see below). But in addition, we point out the fact the *effective* length over which adjacent CNT transfer stress in shear is, probably, much shorter than the size of the macromolecule, and thus, alternative descriptors might be more suitable to predict tensile properties.

In addition to the shape and size of the constituent molecules, CNT fibre properties are strongly determined by the architecture of the collective CNT ensemble. Characterizing the complex hierarchical structure of CNT fibres has been challenging, often requiring multiple techniques. Nevertheless, wide-angle (WAXS) and small-angle (SAXS) X-ray scattering has been proven to capture the key structural features that affect longitudinal properties of CNT fibres. Some of the possibilities of WAXS/SAXS for characterization of CNT fibres are the possibility to: probe both the crystalline and amorphous/porous structure, measure individual CNT filaments, obtain information about the fibre volume rather than the surface, and its compatibility with *in situ* tensile studies.

A brief review of how WAXS/SAXS studies have gone hand in hand with the development of traditional high-performance fibres helps put in context the current level of engineering of CNT fibres and recognize the prospects for further improvement.

X-ray studies were critical in the early development of UHMWPE fibres by providing insights into the fibre structure at the stages of gel spinning, drawing, and annealing. They particularly helped recognize the importance of polymer chain interactions at the point of spinning, hence the introduction of gel-spinning, in order to facilitate polyethylene chain extension and orientation during drawing, leading ultimately to a macromolecular structure with high degree of orientation and crystallinity. A comparison of WAXS patterns for PE at draw ratios spanning from 0 to 31.7 captures very clearly the development of such structure (Figure 12). For highly-drawn fibres the sharp reflections in the WAXS pattern look like those of a single crystal of PE, and the lack of scattering at the small-angle region evidences a rather continuous crystalline structure (Figure 12c). Aramid and PBO fibres spun from liquid crystalline phases have similarly been developed to have a very high level of molecular orientation along the fibre axis (Figure 12 d-f). The strong equatorial features and the well-defined, off-axis, first order (*hkl*) reflections indicate that the molecules are highly aligned and are packed into three-dimensionally (3D) ordered crystallites.

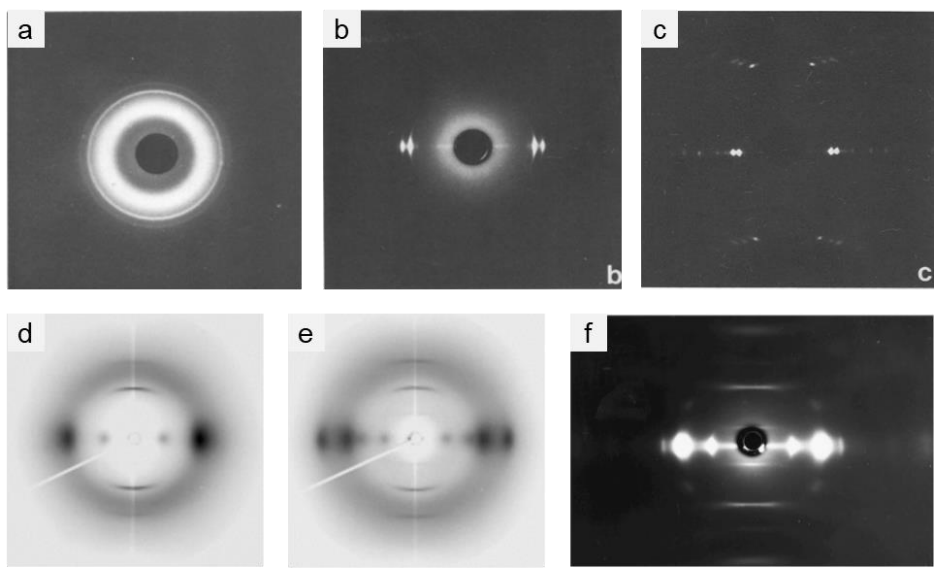

**Figure 12**. WAXS-patterns of gel-spun PE fibres produced with different draw ratios: a) without drawing, b) with draw ratio 15.7; and c) 31.7, respectively. PE fibre drawn at 31.7 exhibited the highest tensile properties (images adapted with permission from [64]. Copyright 1980 Springer Nature), WAXS patterns for PBO fibres d,e) before and after passing through the coagulation water bath (Courtesy of Brookhaven National Laboratory [104]. Copyright 2001 Brookhaven National Laboratory) and f) ready-to-use fibre (adapted with permission from [105]. Copyright Elsevier Science Ltd.).

In the case of CF, crystallite alignment induced by the application of tensile stress at different fabrication stages has been similarly studied by X-ray methods. WAXS/SAXS measurements are instructive to analyse the various different grades of CF (Figure 13). The notion that the structure of high-performance fibres approaches elongated single-crystals is also true for mesophase pitch (MPP) CF (Figure 13a), with (*hkl*) reflections indicating substantial crystallographic registry between graphitic layers and tensile modulus approaching 1 TPa. However, PAN-based fibres which have a much lower degree of graphitization (Figure 13b) are in fact stronger, precisely because of a (defective) microstructure of covalently linked smaller crystalline domains. Nevertheless, for each precursor type, direct correlations between longitudinal mechanical and transport properties with the size of crystalline domains have been found. Table 1 summarizes the *d*-spacing, apparent crystallite sizes $L_c$ and $L_a$, and the full width at half maximum (FWHM) of the *(002)* azimuthal profiles. The table includes modern PAN-based and mesophase pitch-based CF, as well as a historical example WYB carbon fibres produced by the Union Carbide Corporation from rayon precursor in 1960s, with the tensile strength and stiffness of only 0.6 GPa and 41 GPa respectively, due to an irregular cross-sectional shape and low alignment (FWHM of 69°) of small crystalline domains ($L_c$ of 1.8 nm) [21].

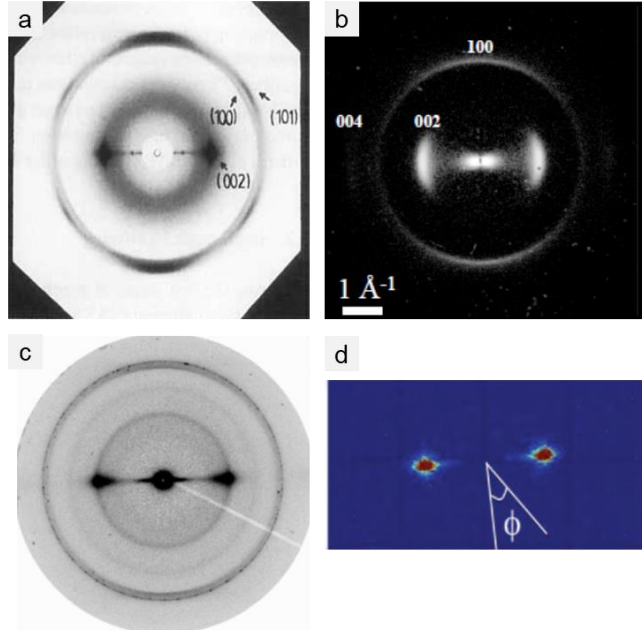

**Figure 13.** WAXS patterns of high-performance fibres: a) mesophase pitch based carbon fibres P120 (reproduced with permission from [106]. Copyright 1990 WILEY-VCH Verlag GmbH & Co.); b) PAN-based carbon fibre HTA7 (adapted from [107] with permission of the International Union of Crystallography http://journals.iucr.org/. Copyright 2000), c) FCCVD-spun CNT fibres showing high alignment but also a powder ring from residual catalyst and carbonaceous impurities (adapted from [108] with permission of the International Union of Crystallography http://journals.iucr.org/. Copyright 2009); and d) LC-spun CNT fibres with very high degree of alignment (adapted with permission from [23]. Copyright 2017 American Chemical Society).

**Table 1.** The structural parameters and related physical properties of CNT fibres compared with some typical carbon fibres.

| Fibre (precursor) | $d_{002}$, nm | $L_c$, nm | $L_a$, nm | FWHM, ° | Modulus, GPA | Strength, GPA |
|---|---|---|---|---|---|---|
| WYB (rayon) [21] | 0.342 | 1.8 | - | 69 | 41 | 0.6 |
| T300 (PAN) [109] | 0.3579 | 1.67 | 4.37 | 27.74 | 231 | 3.2 |
| HTA7 (PAN) [110] | 0.354 | 1.47 | 2.83 | 38 | 198 | 2.7 |
| FT500 (MPP) [110] | 0.343 | 10.06 | 10.22 | 6.7 | 380 | 2.5 |
| P120 (MPP) [111] | 0.3378 | 22.8 | - | 5.6 | 823 | 2.2 |
| K1100 (MPP) [112] | 0.337 | 50 | 40 | 5 | 930 | 3.2 |
| CNT mat (CVD arrays) [113] | 0.342 | - | - | 19.20 | 25 | 0.2 |
| CNT mat (FCCVD, 50% stretched) [70] | - | - | - | 19.3 | 16.2 | 0.38 |
| CNT fibre (FCCVD) [108] | 0.346 | - | - | 12-22 | 50 | 1.1 |
| CNT fibre (FCCVD) [114] | 0.345 | - | ≈ 4 | 18 | 62.4 | 1.1 |

| CNT fibre (LC-spun) [23] | - | - | - | 6.3 | 250 | 2.5 |

There have been comparatively fewer WAXS/SAXS studies on CNT fibres, possibly due to their intrinsically lower degree of crystallinity compared to traditional high-performance fibres, which can difficult data analysis and generally produces such weak scattering intensity that single fibres can often only be analysed at synchrotron facilities, hence WAXS/SAXS patterns are scarce in the literature. Figure 13 c,d shows examples of WAXS patterns for fibres produced by FCCVD and LC-spinning, respectively. The former has moderate degree of alignment, with strong evidence of residual catalyst and carbonaceous impurities observed as powder rings, and abundant elongated pores giving rise to the SAXS feature. The latter one exhibits a pattern that consists essentially of two sharp equatorial reflections from CNT bundles. These large differences in fibre structure and the above-mentioned differences in the envelope of tensile properties reinforce the view that these two fibres are ultimately different grades of CNTF.

A common, interesting feature of CNT fibres is that their degree of molecular perfection, i.e. de degree of conjugation or aromaticity of the nanotubes, is largely independent from the degree of order in their supramolecular ensembles, i.e. the crystalline bundles. This is not the case for CF, in which both are inherently convoluted and develop during the synthesis. Paradoxically, CNT fibres are made up of extremely graphitic molecular constituents, organized in extremely poorly graphitic ensembles. This helps explain why the rows for CNT fibres in Table 1 are large unpopulated and include only approximate values. From this perspective, CNTF could be considered a molecular carbon fibre.

A key recent development has been the use of the uniform stress transfer model (USM) to relate tensile properties to the degree of alignment of the nanotubes in different fibres. Successfully applied to all high-performance polymer fibres (e.g. aramid and rigid-rod) as well as CF, the model reduces the complex fibre to a network of fibrils which can deform elastically by stretching and shear deformation (Figure 14 a,b). Thus, it successfully relates the fibre modulus to the second moment of the orientation distribution function (ODF) for both FCCVD-spun and LC-spun CNT fibres with different degrees of orientation (Figure 14 c). The ODF is directly extracted from WAXS/SAXS data, although Raman spectroscopy [115] and SEM methods have also been proposed [116,117].

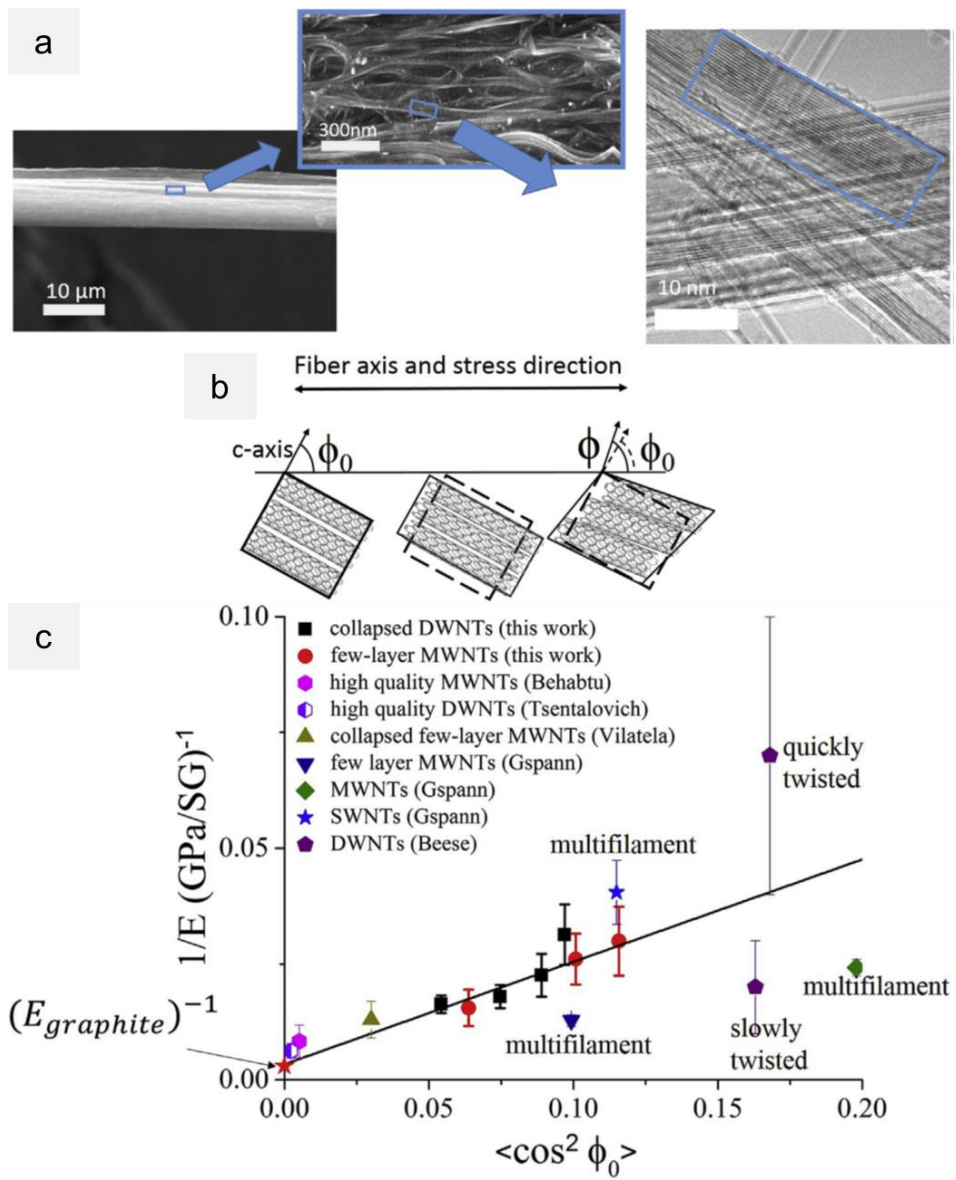

**Figure 14.** a) Electron micrographs of a CNT fibre showing a fibrillar microstructure made up of close packed bundles of nanotubes and b) schematic representation of a single CNT bundle and its contribution to the macroscopic deformation of the fibre: axial stretching by tensile deformation of nanotubes and CNT bundle rotation by shear between nanotube layers. c) Compliance ($1/E_f$) of CNT fibres plotted against the orientation parameter ($<cos^2\phi_0>$) for both in-house fibres and data obtained from the literature. These last correspond to both single [1,23,94] and multifilament samples [118]. Data for twisted CNT yarns is included for reference [119] (images reproduced with permission from [27]. Copyright 2019 Elsevier Ltd.).

The breakthrough of this work is in reducing the complex hierarchical CNT fibre structure to a simpler system with adequate descriptors (i.e., ODF) that can be used in conjunctions with a well-established model, enabling not only comparison of different samples, but also providing accurate predictions of further improvements attainable by maximized alignment. Such emerging use of mechanical models put our understanding of CNT fibre properties on firmer footing and should contribute to the rational design and fabrication of CNT fibres with different combinations of mechanical properties.

*2.4 Post-spinning methods*

The plethora of physical and chemical post-spin treatments suggested for CNT fibres generally aim at improving stress transfer between bundles through better packing, improved alignment, increased interfacial shear strength/modulus, or a combination of them. LC-spun CNT fibres inherently possess relatively dense structure of packed nanotubes with little porosity [22] (Figure 15). In contrast, the structure of dry-spun CNT fibres is highly porous, as they are essentially spun from aerogels. In the fibres spun from vertical CNT arrays, "air pockets" can occupy for 40% of the total volume of a fibre [120]. For FCCVD-spun fibres, the initial porosity can exceed 50% depending on the synthesis conditions [7].

Liquid infiltration and densification by solvents and polymers has been widely considered a successful technique to increase the packing efficiency of bundles and enhance the structural coherence of CNT fibres and mats. Kanakaraj *et al.* modified the spinning line and densification process by extruding the array-spun ribbon through a Teflon die first, then guiding it through a heated solvent bath (N-Methyl-2-pyrrolidone (NMP) or acetone) under some tension prior to winding on a bobbin [120]. Although the role of pre-stretching on increasing the alignment of bundles within a fibre has not been evaluated, this example demonstrates an increase in density and mechanical performance. The NMP stretch-densified fibres exhibited much less porosity (4% *versus* 46% for the pristine fibres spun only with twisting applied (Figure 16 a-d) and the tensile strength doubled that of the pristine fibres (1.2 GPa *versus* 0.6 GPa).

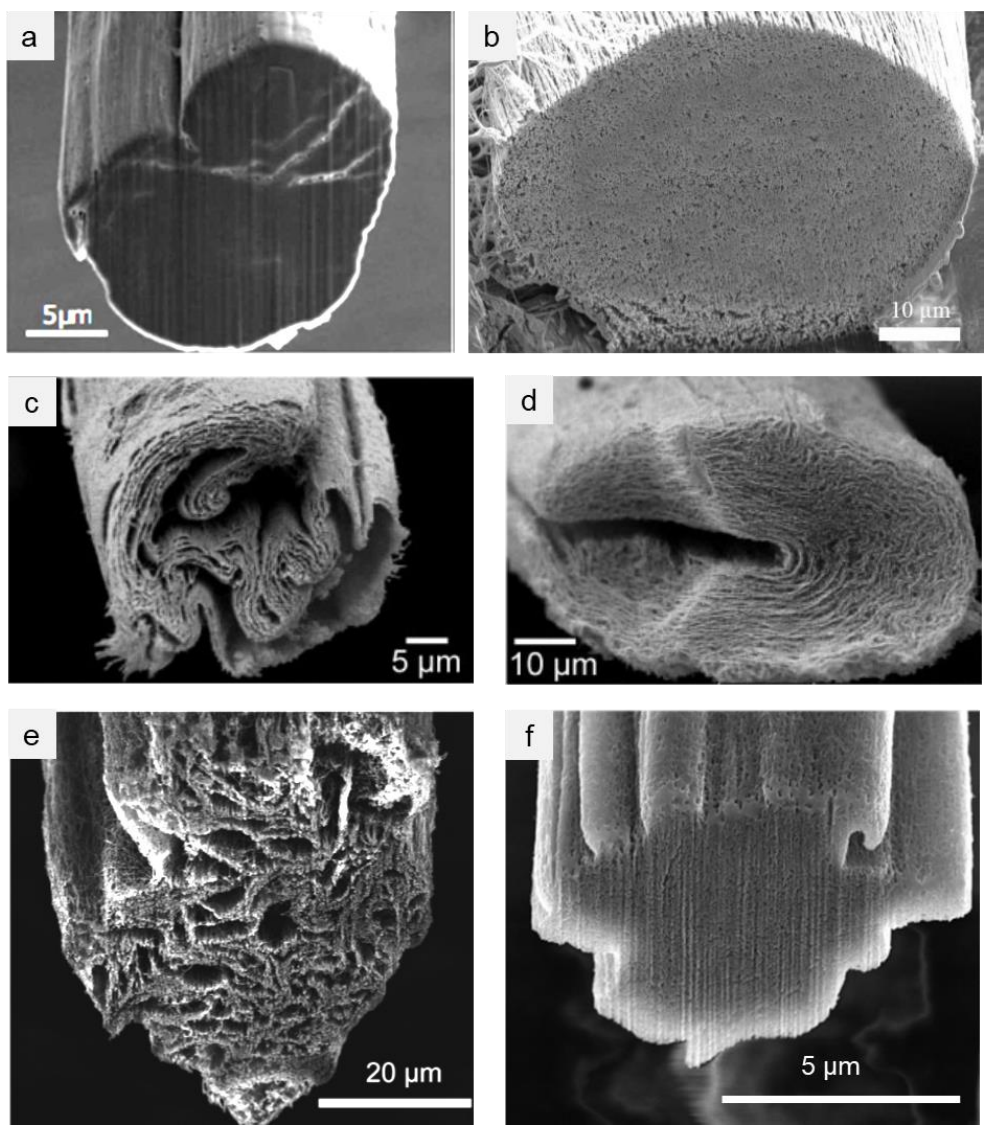

**Figure 15.** SEM images of the Focused Ion Beam-milled cross-sections of CNT fibres spun via different methods: a) wet-spinning from LC solutions (adapted with permission from [23]. Copyright 2017 American Chemical Society); b) dry-spinning from vertical CNT arrays (adapted with permission from [120]. Copyright 2018 IOP Publishing Ltd.); c-f) dry-spinning via FCCVD, produced by different groups at different synthesis conditions, (images c,d adapted with permission from [11]. Copyright 2010 WILEY-VCH Verlag GmbH & Co.), (image e adapted with permission from [7]. Copyright 2014 Royal Society of Chemistry), (image f adapted with permission from [118]. Copyright 2017 Elsevier Ltd.).

Cho *et al.* applied wet-stretching in NMP and chloro sulfonic acid (CSA) to densify the structure and reduce the inter-bundle pores of the CNT fibre spun by the FCCVD method (Figure 16 e-h) [121]. Interestingly, the authors have observed that simple addition of NMP increased the tensile strength of a fibre from 0.22 GPA to 0.72 GPa, while further wet-stretching (up to 7%) did not show any additional effect on the mechanical properties. In contrast, wet-stretching in CSA increased tensile properties in a proportion to applied strain, with the tensile strength reached 1.61 GPa for a fibre wet-stretched in CSA to 15%.

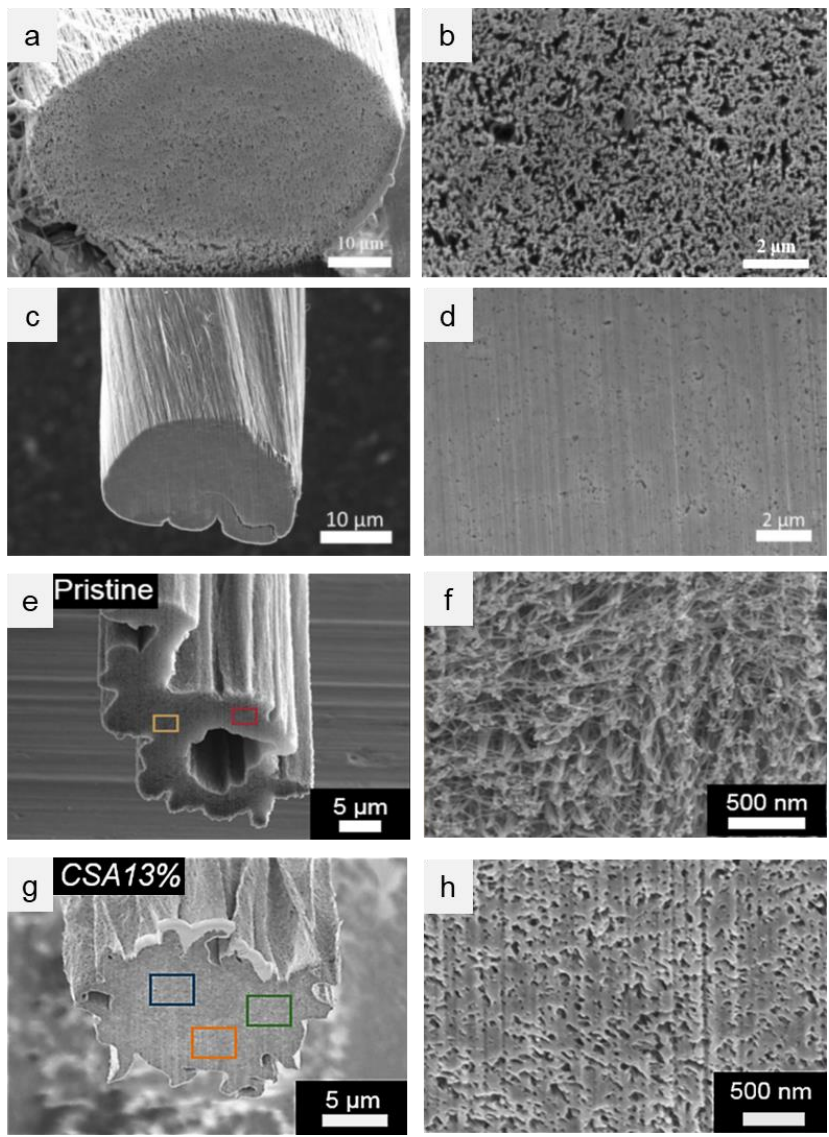

**Figure 16.** SEM images of the Focused Ion Beam cross-sections of as-synthesized and wet-stretched CNT fibres with significantly improved packing of nanotube bundles and reduced porosity: a,b) pristine CNT fibre dry-spun from vertical arrays (images adapted with permission from [120]. Copyright 2018 IOP Publishing Ltd.); c,d) the same fibre stretched-densified in NMP (images adapted with permission from [120]. Copyright 2018 IOP Publishing Ltd.); e,f) pristine CNT fibre dry-spun via FCCVD (images adapted from [121]. Copyright 2018 Elsevier Ltd.) and

g,h) the same fibre after stretching in CSA for 13% (images adapted with permission from [121]. Copyright 2018 Elsevier Ltd.).

These examples demonstrate the importance of post-spin stretching, considered by the present authors as an essential processing step of CNT fibres. Precisely for this reason, future work should aim at a better characterization of the structure of CNT fibres in this "gel" state and understanding the physics of the post-spin wet-drawing process. As a corollary, it would be of interest to clarify and quantify the effects of increasing density through better CNT packing on tensile properties, noting that graphite interlayer potentials fall very rapidly with separation.

Additional post-spinning treatments, such as UV [122], irradiation [123-125], infiltration with polymeric compounds and molecular couplings and cross-linking [29,126-131] can enhance the lateral interactions between nanotubes or their bundles. Further work should be directed at developing metrics to compare these treatments and assess their applicability to different CNT fibres, for example, by quantifying their effect on effective shear length and stress transfer length. An interesting aspect to consider is combining of these post-spinning treatments with stretching, since it is possible that their improvement in mechanical properties is additive, as they target different CNTF features: interfacial stress transfer and CNT spatial arrangement, respectively.

**PART 3. Properties of CNT fibre arrays and their composites**

*3.1 Manufacturing of CNT fibre-reinforced composites: polymer infusion and volume fraction limitations*

The general procedure to make CNT fibre composites is to infiltrate CNT preforms (stacks of CNT sheets, fibres, or thin mats) with a polymer matrix by methods such as: casting a vacuum-degassed liquid resin over the preforms [116,132,133], spraying [134] or immersion of preforms in the diluted polymer solutions [135-139]. Facile polymer penetration into CNT preforms has been demonstrated when using low-viscosity epoxy solutions of various concentrations in acetone [135,136,139], usually followed by drying in a vacuum oven to remove the solvent. The infiltrated CNT fibres and mats are then placed in the home-made customized moulds: silicone [132] or metal [116]), in-between of glass blocks [135] or Teflon sheets [132], or even cured in an oven as free-standing samples [136] (Figure 17 a,b). In some cases, a second vacuum degassing step or curing in a vacuum oven were applied for the infiltrated CNT materials [116,132,133] to minimize potential voids.

A similar approach of using the dilute solutions to promote polymer infiltration and tailor the resultant polymer matrix content in the composites has been utilized for polyimide matrix 0.1% wt. solution in NMP [134], and for various thermoplastic polymers as polyvinyl alcohol (1%wt. solution in DI water) [137], nylon 6,6 (1%wt. solution in phenol) [138], and for thermoplastic polyurethane and polystyrene dissolved in dichloromethane [140]. Although almost all literature sources claim complete polymer infiltration of CNT material in composite manufacturing, there is a clear lack of data on porosity of the resultant composites, with limited examples of the control over the density and voids content [116,136,139,140]. In these reports the CNT arrays are usually thin and the composites produced by methods that favour infiltration but differ substantially from those used on the scale of a laminated structural composites.

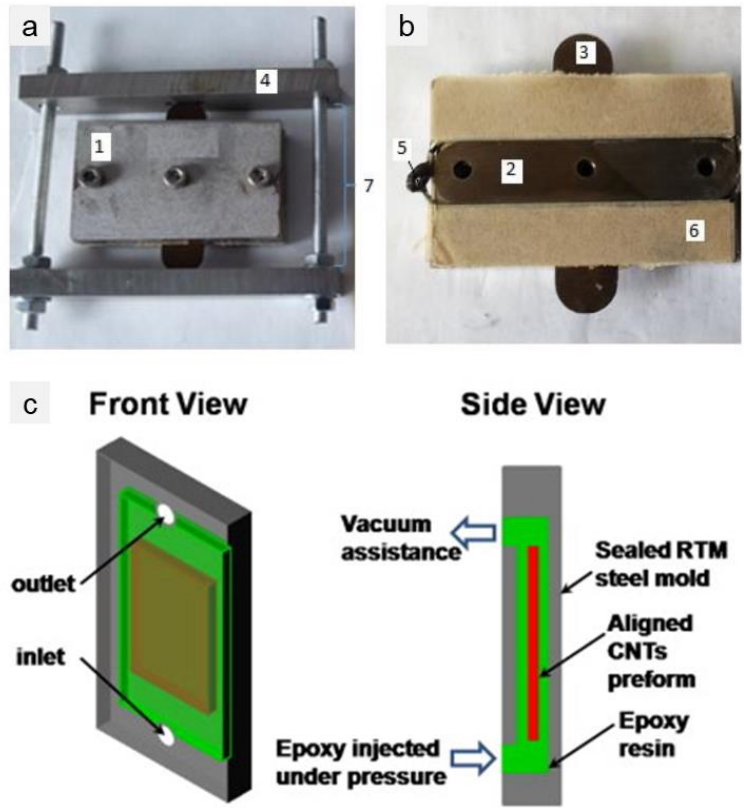

**Figure 17.** Examples of customized moulds for manufacturing the CNT fibre reinforced composites a,b) The sectional mould (1-3), designed to control the applied pressure by controlling the distance between the plates (7) in the clamping frame (4), with a sample (5) placed with anti-adhesive paper (6) (images courtesy by A. Mikhalchan); c) A Schematic of the RTM mould with a CNT preform, while the thickness of the resultant composites is controlled by inserting steel strips of various thicknesses between two parts of the mould (images adapted with permission from [141]. Copyright 2010 Elsevier Ltd.).

Cheng *et al.* [141] fabricated epoxy composites reinforced with aligned CNT sheets by adapting a RTM process (Figure 17c). They elevated temperature to 60 °C in order to keep viscosity of the epoxy system as low as 50 cPa to overflow and infiltrate the CNT preforms at the applied pressure of 0.2 MPa. The authors concluded that composites with volume fraction of nanotubes of 20% cannot be fabricated by the RTM process because of low permeability of CNT preforms, and their attempt to make a composite with 6000 CNT sheets (with an expected thickness of >300 μm) failed. Recent work in our laboratory using vacuum bag resin infusion of hybrid CF preforms with a thin ≈10 μm layer of CNT interleaf in a 30 x 30 cm panel, showed that complete infiltration requires using two layers of distribution media and longer infusion time, as examples of some of the changes relative to the standard fabrication process in order to ensure adequate infusion.

The autoclave curing process has so far only been applied to CNT preforms hybridized with conventional CF prepregs. In the autoclave curing of CNT/CF hybrids, the resin from the CF prepregs is forced to flow in the dry CNT interlayers predominantly in the through-thickness direction. The through-thickness permeability of CNT sheets is reported at about $10^{-19} - 10^{-17}$ m$^2$ [142] which is much lower than that of conventional CF materials, e.g. $10^{-13}$ m$^2$ for unidirectional CF fabric of 231 g/m$^2$ areal density [143] or $10^{-17}$ m$^2$ measured for commercial UD T700 prepregs with 34 wt.% resin content and 200 g/m$^2$ areal weigh with the stacking sequence of the [+45/45]$_S$ and [0/90]$_S$ [144]. The examples of CF/CNT hybrids confirm that autoclave curing conditions need to be optimized to allow longer period for the resin to flow through dry CNT interleaves [145].

The origin of voids itself in the CNT fibre-reinforced composites is unclear. Whereas in classical composites voids emerge at the fibre/matrix interface region, in composites with CNT materials, the voids are usually detected within the CNT assembly. The nanoscale voids could appear due to resin starvation if the volume fraction of nanotubes is high, as observed in composites reinforced with CNT fibres and mats [116, 136]. Nanoscale voids have been detected in the CNT fibres, mainly at the junctions of CNT bundles, in autoclave-cured CF/CNTF hybrid laminates with the epoxy matrix available from the adjacent CF prepreg layers [146]. One possible reason is that the resin content in the prepreg might not be sufficient to fill the CNTF pores, which can represent over 50% of the CNTF sample volume.

*3.2 Making prepregs with CNT fabrics for composite manufacturing*

The shift towards better control over the complete infiltration and volume fraction, as well as for scaling the size of the laminates is possible with introduction of CNT fabric-based prepregs in composite manufacturing. This is another sphere where a significant progress is anticipated in the nearest future. The first examples introduced the fabrication of prepregs by using the commercially available epoxy films. Ogasawara *et al.* hot-pressed 20-100 array-spun CNT sheets with epoxy film for 3 min at 90° C on the polytetrafluoroethylene (PTFE) support sheet, to obtain a CNT prepreg of 24-33 μm thickness with good drapability and tackiness [147]. Later, Nam *et al.* pressed 100 stacked array-spun CNT sheets with epoxy film for 5 minutes at 100 °C and used the prepregs in composite manufacturing by hot-press [148]. Wang *et al.* [145] adopted the process for making the prepregs with CNT mats synthesized via FCCVD; however, only used them for hybrid composites. Jiang *et al.* [134] produced prepregs by infiltrating CNT sheets with 0.1% wt. polyimide/NMP solution, the prepared prepreg of 10-15 μm thickness was dried in a vacuum oven to remove NMP and then hot-pressed in a vacuum oven to form a composite with completed imidization. Nguen *et al.* [149] prepared the prepregs by soaking the 25 μm thickness commercial CNT fabrics in a diluted BMI/acetone solution for 15min followed by drying at room temperature for 24 h, and then used them for hybridizing with CF prepregs and out-of-autoclave curing by electric current; however, the authors did not disclose further information on the prepregs or composite mechanical properties.

These sparse examples demonstrate the possibility of making the prepregs on the basis of CNT fabrics. Further detailed analysis is necessary to optimize the manufacturing process and evaluate the properties of such prepregs in the similar was as of commercially produced prepregs with high-performance fibres. The analysis may include evaluation of drapability, tack, shelf and pot life, the storage conditions, etc. that have not yet been determined. Last but not least, making the prepregs to compete with industrial processing is a question of availability of larger amount of CNT materials.

*3.3 Challenges in testing the CNT fibre-reinforced composites*

Composite manufacturing requires amount of CNTF materials that are challenging to produce at the laboratory-scale. The established standard methods for mechanical testing are thus usually adapted to the dimensions of the composites produced recently with CNT fibres. Compared to standards for fibre-reinforced laminates (e.g., ASTM D3039 or ASTM D7205) involving samples of tens of cm$^2$ and millimeter thickness, CNT-based composites are often substantially smaller, far from standard requirements (Figure 18). According to literature, most of the tensile tests on CNTF composites were done on coupons with total length of 10 mm, tested at the gauge length of 6 mm [134,138]. A few examples of relatively larger samples had a length of only a few centimeters, a few mm width, and thickness of 50-500 microns [116,150]. Even these samples require a substantial CNTF synthesis. Note, that a composite laminate with a size similar to a typical carbon fibre reinforced one as on Figure 18 h (20 by 30 cm) and a volume fraction of 50-60% vol, will require more than 150 grams of CNT fibres as a rough estimation. This is close to the year's CNTF production in a research laboratory under present synthesis conditions.

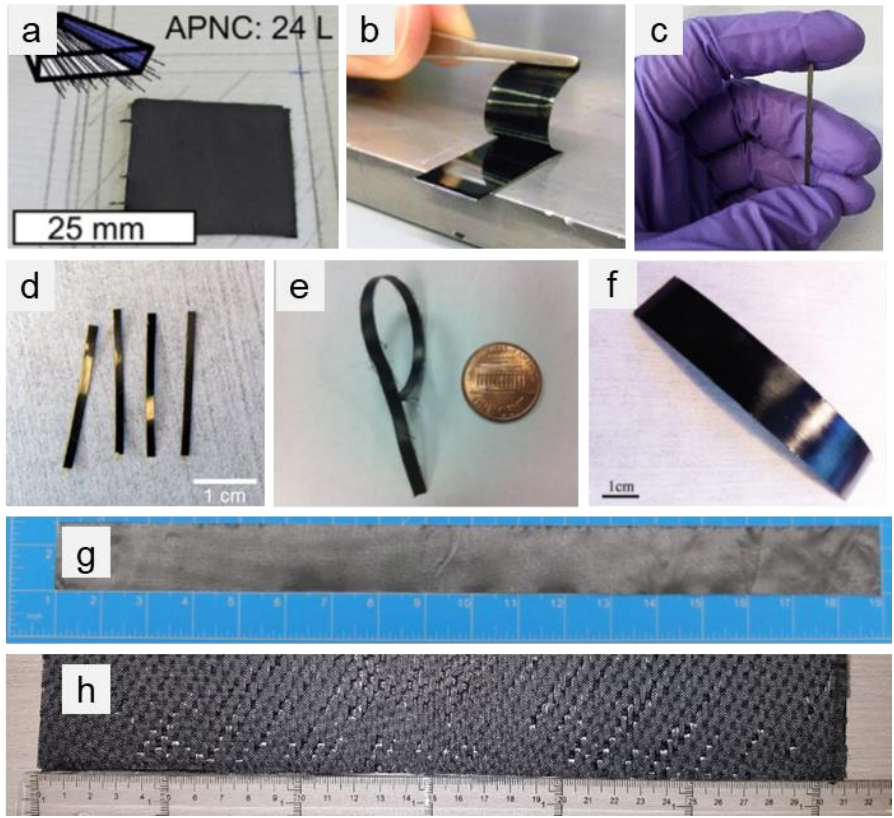

**Figure 18.** Examples of the small composites reinforced with different CNT assemblies: a, b) aligned CNT mats made by "domino-push" from vertical CNT arrays (image a adapted with permission from [150]. Copyright 2015 Elsevier Ltd.), (image b adapted with permission from [135]. Copyright 2010 Elsevier Ltd.); c) epoxy composite reinforced with high volume fraction of aligned CNT fibres (adapted with permission from [116]. Copyright 2016 Springer Nature); d) examples of CNT epoxy composite tensile test coupons (adapted with permission from [135]. Copyright 2010 Elsevier Ltd.); e) CNT/nylon 6,6 composite sample (adapted with permission from [138]. Copyright 2011 Elsevier Ltd.); f) CNT/polyimide prepreg prepared for composite manufacturing (adapted with permission from [134]. Copyright 2016 SAGE Publications); g) example of 20-50 μm thin composite cut from a 30 cm x 45 cm plate made by hot-pressing of commercial CNT mat with poly[(benzoyl-1, 4-phenylene)-*co*-(1, 3-phenylene)] (adapted with permission from [71]. Copyright 2017 Springer Nature); h) a part of a typical carbon fibre/epoxy laminate plate of 20 cm x 30 cm is shown for comparison (image courtesy by A. Mikhalchan).

*3.4 Transferring filament properties to large arrays: fabrics without fibres*

Current specific tensile properties of CNT fibres are already in the high-performance range, making the material attractive for composites. Yet, a longstanding hurdle has been to produce large-area fabrics or similar arrays that efficiently reflect the properties of the individual CNT filaments. Figure 19 shows the longitudinal tensile strength and specific electrical conductivity of a CNT fibre, thin tows of around 50 CNT fibres, thin laboratory-scale fabrics (around 500 – 1000 filaments) and a commercial unidirectional CNT fabric. The data are only very approximate, collected over years at the author's laboratory rather than corresponding to a thorough study on assembly. Nevertheless, they capture the large difference in properties comparing single aerogel filaments with the yarns and fabrics that are then used to make composites.

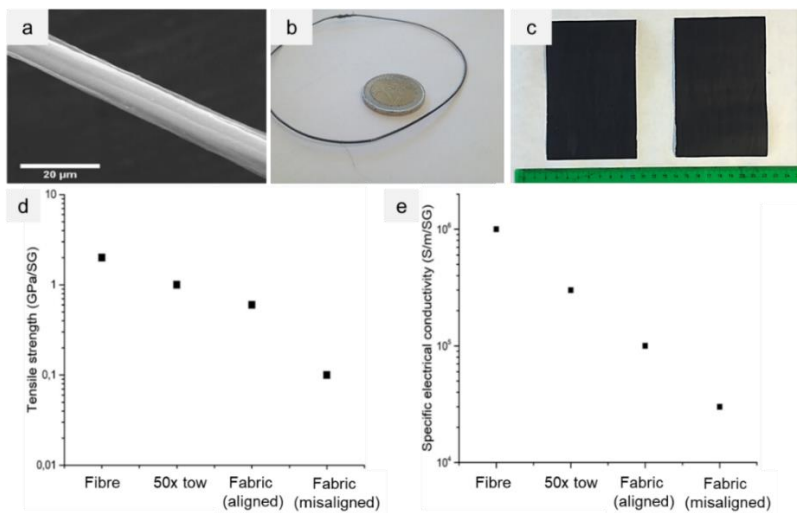

**Figure 19.** a-c) Examples of the CNT assemblies: individual fibre, tow of 50 fibres, and unidirectional multifilament fabrics, d) e) the corresponding longitudinal tensile strength and specific electrical conductivity of these materials.

Such behavior reflects mainly misalignment of filaments in the fabrics, leading to non-cooperative loading when the fabric is stressed. This seems, in essence, a generic textile engineering problem, which could be tackled as scale-up efforts develop multiple parallel fibre spinning and established fabric "straightening" methods can be applied to fabrics. But an interesting feature of dry-spun CNT materials is the possibility to directly merge CNT aerogel into a fabric during spinning. Figure 20 shows examples of production of unidirectional CNT fabrics by overlaying aerogel filaments. In the resulting multifilament fabric, also referred to in literature as mats or sheets, the aerogel filaments bind irreversibly through van der Waals forces and merge into a continuous network.

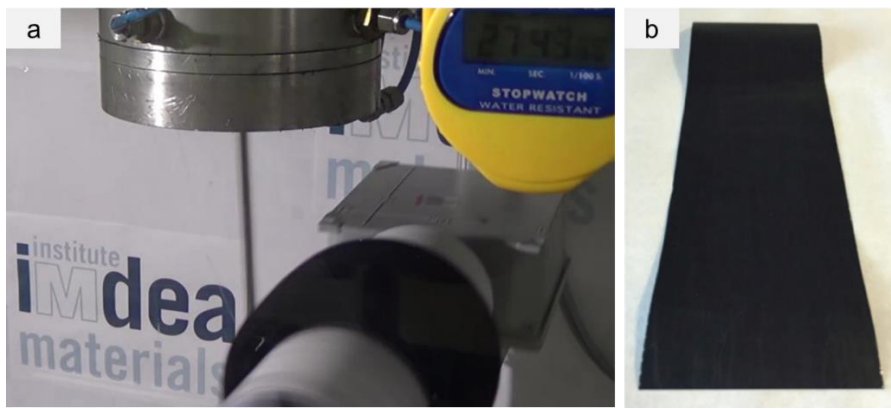

**Figure 20.** Production of unidirectional multifilament CNT fabrics by overlaying CNT aerogel filaments: a) spinning of CNT filaments directly from the gas phase; b) ready-to-use CNT fabric collected on a detachable substrate (adapted with permission from [151]. Copyright 2017 WILEY-VCH Verlag GmbH & Co.).

*3.5 Wet-stretching before composite fabrication*

These fabrics have been the preferred material for use in composites and thus there have been various proposed methods to improve the fabric integration and properties. Some examples of enhancing the alignment of CNT bundles via wet-stretching techniques have been shown for dry-spun CNT mats. Gbordzoe *et al.* [113] stretched CNT mats spun from vertical arrays in acetone to 1.0%. According to ODF profiles of X-ray scattering analysis, FWHM decreased from about 19° to about 16° (Figure 21 a-c) and the tensile strength increased by 2.5 times to about 500 MPa. However, further stretching of mats to 1.5% led to reduction of their electrical and mechanical performance due to breakage and slippage of bundles.

Liang *et al.* have stretched mats of randomly-aligned CNT synthesized via FCCVD up to 40% in a dry state in order to improve their mechanical performance, while attempts to stretch over this level were not successful [152]. The similar mats have been stretched to remarkable 50% [70] and 65% [71] with the aid of polymer molecules (epoxy resin), which provided support for the re-arranging of CNT bundles under tension. After removing the resin with acetone, the stretched mats remained stable and exhibited good integrity with drastically enhanced alignment of CNT bundles. According to WAXS analysis, the FWHM decreased from 52.1° for the 15%-stretched sample to 19.3° for 50%-stretched [70] and 14° for 65%-stretched one [71] (Figure 21 d-f). Apart from reduced waviness of CNT bundles in the load direction, the authors have observed the self-assembling and denser packing of the bundles, which altogether benefited to the load carrying ability of the fibres and their mechanical performance. The tensile strength and Young's modulus increased from about 120 MPa and 1.5GPa for the initial mat to 423MPa strength and 21.6GPa for the 65%-stretched mat.

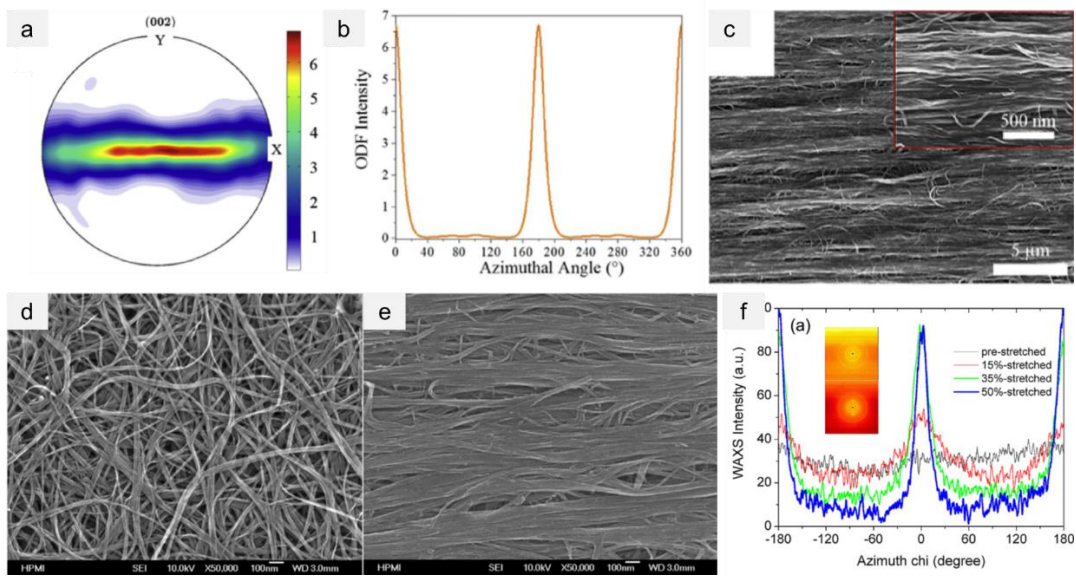

**Figure 21.** a) the Orientation Distribution Function (ODF) pole texture analysis with b) corresponding ODF profile reconstructed for c) the array-spun CNT mat stretched in acetone to 1%, SEM image (inset: higher magnification) (images a-c adapted with permission from [113]. Copyright 2017 Elsevier Ltd.). d) and e) SEM images of the pristine and 50%-stretched CNT mat from FCCVD method, with f) integrated 2D X-ray scattering intensity (summed over intervals 24<2θ<30) displaying the azimuth peaks for nanotubes in the differently stretched mats (inset: 2D X-ray scattering patterns for initial (up) and 50%-stretched (bottom) CNT mat) (images d-f adapted with permission from [70]. Copyright 2016 WILEY-VCH Verlag GmbH & Co.).

Wang *et al.* [153] stretched randomly-aligned FCCVD-spun CNT mats by 40% in ethanol-water solution (weight ratio 1:1), followed by heating it at 350 °C for 1 hour under tension. After wet-stretching and post-heating, the CNT mats displayed notable improvement in tensile properties with the strength and stiffness enhanced from 186 MPa and 3.2 GPa to 1.17 GPa and 127.1 GPa, respectively. According to the authors, wet-stretching allows to achieve large extent rearrangement of nanotubes with the aid of lubrication from the solvent, whereas post-stretching heating removes the CNT surface byproducts for stronger tube-tube interactions. Although the mechanisms behind the stretching processes are not yet fully understood, these results demonstrate the positive effect of stretching on the maximizing alignment of the CNT assemblies.

*3.6 Tensile properties of CNT fibre composites*

Figure 22 summarizes data on the tensile properties of thermoplastic and thermoset composites reinforced with CNT assemblies (e.g., vertical arrays, fibres, mats, papers). Early work in the field using pristine unidirectional fabrics produced moderate tensile strengths and moduli [116,132,133,135-137,139-141,150,154], but nevertheless, exposed some of the fundamental properties of these materials, for instance, the substantial improvement in fabric properties upon polymer infiltration.

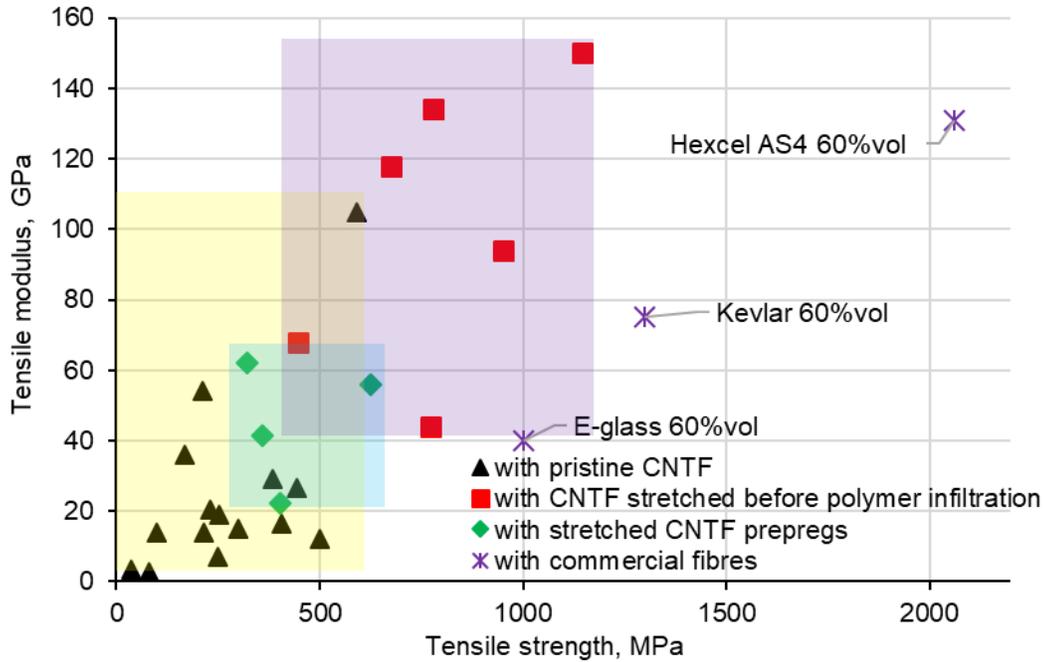

**Figure 22**. Strength and stiffness of composites reinforced with different CNT assemblies showing the potential of stretching the pristine CNT materials or the prepregs to maximize mechanical performance.

Extensive work on composites with dispersed nanotubes as fillers had shown previously that slight waviness of individualized nanotubes significantly reduces the effective reinforcing potential compared to straight nanotubes [155-159]. A similar behavior can be expected for composites reinforced with high volume fraction of aggregated nanotubes in the form of fibres. Subsequent efforts, thus, focused on pre-stretching of CNT fabrics using the methods discussed above prior to polymer matrix infiltration, and then integrating them in composites.

Data for composites with pre-stretched CNT fabrics included in Figure 22 show indeed a very large improvement in longitudinal strength and modulus. 3% stretching of the array-spun CNT mats before resin infiltration resulted in the tensile strength of 448 MPa and modulus of 67.9 GPa in comparison with 211 MPa and 54.0 GPa for the composite reinforced with pristine CNT mats [154]. Liang *et al.* reinforced thermoplastic composites with highly-stretched FCCVD CNT mats and showed that both tensile strength and stiffness increased as a consequence of better alignment of CNT bundles. The composites manufactured with randomly-oriented nanotubes exhibited the tensile strength and modulus of 384 MPa and 29 GPa, respectively, which increased drastically to 950 MPa and 94 GPa for the composites reinforced with 50%-stretched CNT mats [70], and to 1145 MPa and 150 GPa for the composites with 65%-stretched CNT mats [71]. Yu *et al.* [137] used water-assisted shear stretching of random CNT mats prior to their infiltration with polyvinyl alcohol (PVA). Due to increased orientation of CNT bundles, tensile strength and stiffness of the CNT/PVA composites (about 80% wt CNT fraction) increased from 248 MPa and 7 GPa for the composite reinforced with a random CNT mat to 773 MPa and 44 GPa for the composite manufactured with the stretched mat. Nam *et.al.* enhanced alignment of arrays-spun CNT sheets before embedding them into a polymer matrix through the technique similar to [148], consisting of drawing, stretching, winding, and pressing steps [160]. The epoxy composites reinforced with stretch-drawn and winded CNT sheets showed around 675 MPa strength and 118 GPa modulus, and the samples reinforced with stretch-drawn and press-winded CNT sheets exhibited remarkable 780 MPa strength and 134 GPa modulus. Note that the properties of these pre-stretched composites are in most case far superior to those of the dry fabric before and after stretching, a consequence of both improved alignment and stress transfer through the polymer matrix.

One more possibility to enhance CNT alignment in the composites is stretching directly CNT prepregs, i.e. preforms or fabrics already infused with thermoplastic or thermosetting polymers, which also seem attractive from the point of view of manufacturing. Several works on thin (10-20 micron) stretched CNT prepregs have reported notable improvements in mechanical properties. Resin infused CNT preforms made of vertically-aligned nanotubes were strained to 5% before curing [135]. This led to the strength and modulus enhanced from 300 MPa and 15 GPa for the initial composite to 402 MPa and 22.3 GPa for the composite reinforced with pre-stretched prepregs (27 %vol fraction). In another example [154], hot stretching of array-spun CNT mats/epoxy prepregs to 5 % enhanced the tensile strength by 50% and stiffness by 20%, so they reached about 320 MPa and 62 GPa respectively. Jiang *et al.* [134] applied low velocity stretching of 1 to 6% to array-spun CNT mats/polyimide prepregs and observed the tensile strength and stiffness increased to 358 MPa and 41.3 GPa compared to the composite reinforced with the pristine CNT mat (98.9 MPa and 14 GPa, respectively). Wang *et al.* [138] applied hot stretching for array-spun CNT/nylon 6,6 composites. After stretching to 6%, the tensile strength and stiffness enhanced from 215 MPa and 14 GPa for the initial composite to 625 MPa and 56 GPa for the stretched one.

As seen from Figure 22 above, stretching of CNT materials before manufacturing the composites results in a much larger enhancement of mechanical performance compared to stretching of CNT-based prepregs; although we note the difficulty in comparing these literature data because of differences in fibre volume fraction, thickness, type of CNT fabric and type of polymer matrix, amongst other composite parameters. In respect to polymer matrixes used, most of the works were done with epoxy resin, with few examples of thermoplastic-based composites (mentioned above). Although, the longitudinal modulus and strength of the composites reinforced with CNT fibres could depend slightly on the matrix chosen, they are dominated by the CNT fibre properties and alignment, as in the case for composites with conventional fibres. The correlation of the *specific* modulus of the polymer-infiltrated CNT fibres with the polymer matrix Young's modulus and unusual aspects of their mechanics are discussed in the following section 3.8. It is notable from Figure 22, however, that the highest improvement of the mechanical performance approaching the range of 1 GPa strength and 100-150 GPa modulus has been reached for composites with epoxy (typically 100 MPa strength and 2-3 GPa stiffness) and LC polyethersulfone-based (around 250 MPa strength and 8 GPa stiffness [161]) matrixes when they were reinforced with CNT ensembles with maximized alignment of CNT bundles. It is at present unclear why solvent-stretching is more effective than polymer-stretching. But from the emerging results and the lessons from gel-polymer fibre drawing, it is evident that a deeper understanding of the CNT fibre/fabric structure in the "gel" state (i.e., aerogel in FCCVD synthesis) and the final structure and tensile properties of the composite material are essential. Raman spectroscopy and WAXS/SAXS of infused samples under stretching are likely to provide key insights into molecular stress transfer and into the evolution of the porous crystalline network structure of the fibres.

A comparison with traditional composite materials shows that composites reinforced with CNT fibres and mats already have longitudinal properties in the high-performance range. They outperform commercial unidirectional laminates reinforced with glass fibres, especially, in terms of the density, yet cannot match with composites reinforced with para-aramid or carbon fibres. This is encouraging considering the large room for fibre/fabric development discussed above, as well as expected increases in fibre volume fraction once composite fabrication is optimized.

*3.7 Towards the determination of CNT fibre lamina properties*

The longitudinal properties of CNT fibre-reinforced unidirectional composites are reasonably well established, but there is a general lack of data on properties in transverse direction and under compression, both critical for the mechanical properties of laminate composites. This is partly a consequence of the aforementioned limitations on sample size/number and partially of the lack of information on transverse/compressive properties of CNT fibres themselves.

Zu *et al.* [162] estimated the axial compressive strength of single CNT fibres from a tensile recoil analysis according to the method developed by Allen [163]. The key assumption there was no energy dissipation and that the magnitude of the compressive stress wave generated during a specimen recoil was equal in magnitude to but of opposite sign to the initial tensile stress. The recoil compressive strength of a pristine CNT fibre spun from vertical CVD arrays was determined of 416 MPa, higher than that of Kevlar-49 fibre (365 MPa) [163] and comparable to that of pitch-based Thornel P130 carbon fibre (410 MPa) [164] obtained in similar tensile recoil measurements. Epoxy infiltrated CNT fibre showed a pronounced increase in the recoil compressive strength up to a value of 573 MPa. Interestingly, the previous analysis of pristine CNT fibres with moderate tensile properties spun via FCCVD revealed the lower recoil compressive strength in the range of 171-177 MPa [67], which the authors ascribed to the structural difference resulting from these two fibre spinning methods.

Gao *et al.* [165] monitored axial compressive stresses through micro-Raman measurements in epoxy-embedded CVD-array spun CNT composite fibres under thermally-induced stresses. Evidence of CNT buckling was not found until axial stresses reached around 2.3GPa, and the fibre could be further stressed to a remarkable 3.5GPa. Calculations on matrix-infiltrated CNT fibres confirm that the critical stress for nanotubes to bend or buckle is in excess of 2GPa compared to 1.8 GPa of pristine fibres [166], indicating that surrounding polymer would supply lateral support to hamper the onset of bending or buckling of nanotubes at relatively low stress.

Li *et al.* [167] analyzed the transverse compressive response of pristine commercial dry-spun CNT fibres (both, CVD-array and FCCVD-spun) through the single fibre transverse compression test, in which a fibre was laid on a flat sapphire platen and compressed by another sapphire platen at quasi-static loading. The transverse compressive moduli measured from the slope of the linear elastic region of the curves were 0.21 GPa for the FCCVD-spun CNT fibre and 1.73 GPa for the CVD-array-spun fibre, and the fibres fractured when the transverse compressive stress reached 0.796 GPa and 1.036 GPa, respectively. According to the authors, the higher transverse compressive performance of the array-spun fibres was attributed to better CNT axial alignment, more densely packed structure, and a larger inter-tube contact area, whereas the less densely packed structure of the FCCVD-spun fibres resulted in lower load transfer efficiency. The data are compared to those of individual conventional high-performance fibres transversely compressed in a similar way between two mirror-like polished flat steel beds (Table 2) [168].

**Table 2.** Strength and stiffness of high-performance fibres in tension and transverse compression

| Fibre | Mechanical properties | | | | | |
| --- | --- | --- | --- | --- | --- | --- |
| | Tension | | Axial compression | | Transverse compression | |
| | σ, GPa | E, GPa | σ, GPa | E, GPa | σ, GPa | E, GPa |
| T300 (PAN) carbon fibre [168,169] | 3.24 | 231 | 2.88 | - | 2.73 | 6.03 |
| P25 (pitch) carbon fibre [168,169] | 1.38 | 159 | 1.15 | - | 0.64 | 9.95 |
| P120 (pitch) carbon fibre [168,169] | 2.24 | 827 | 0.45 | - | 0.079 | 3.08 |
| Kevlar-29 [163,168] | 3.03 | 98.4 | 0.350 | - | 0.056 | 2.59 |
| Kevlar-49 [163,168] | 2.55 | 129.6 | 0.365 | - | 0.060 | 2.49 |
| CNT fibre (CVD array) [162,167,170] | 1.2 | 43.3 | 0.416 | - | 1.036 | 1.73 |
| CNT fibre (FCCVD) [162,167] | - | - | 0.177 | - | 0.796 | 0.21 |

A prevailing question is whether the high inherent flexibility of dry CNT fibres will inevitably lead to poor properties of CNT-based composites in compression. The scarce preliminary studies suggest the contrary. In an early paper [133], small cylindrical specimens were produced by epoxy diffusion into an array of aligned CNT fibres cured in a customized cylindrical mould, each composite sample contained about 100 m of CNT fibre (14% vol fraction). To perform compression testing, the special manipulations with the cylindrical samples of 1.8 mm width and 4 mm length were necessary. They were cut into shorter pieces of about 4 mm long and both ends of the specimens were carefully polished to control that both circular faces to be parallel to each other and perpendicular to the cylinder axis. The obtained results confirm that the infiltration of epoxy into fibres effectively prevents compressive instability of the CNT bundle network, and reinforces the structure against compressive buckling. A yield stress was observed in the vicinity of 130 MPa (118 MPa/SG), after which severe plastic deformation occurred up to strains of 30%. Both compressive modulus and yield stress were close to a rule of mixtures.

Later, the authors adapted the ASTM D7264 standard for three point bend test of the small CNT fibre-reinforced composites with dimensions of 40 × 2 × 0.2–0.8 mm (length x width x thickness) [116]. They kept the specimen length of about 20% longer than the support span distance and preserved the span-to thickness ratio to be near to 40:1 for all tested samples, as recommended by the standard for the case of highly anisotropic materials. The composites reinforced with uncondensed CNT mat (39% vol CNT fraction) exhibited flexural strength and modulus of 185 MPa and 17 GPa, and reinforced with high-strength CNT fibres (30 % vol fraction) showed 514 MPa strength and 47 GPa modulus, correspondingly, both greatly enhanced relative to pure epoxy matrix (124.5 MPa and 2.1 GPa). Buckling as a form of compressive failure in bending was only observed in composites reinforced with CNT mats. The authors concluded that the additional coherence between the CNT bundles and improved quality of their alignment achieved at the densification step in CNT fibres, possibly contributes to their greater resistance to compressive buckling.

The limited data available suggests that the axial compressive strength of CNT fibres are comparable to those of other high-performance fibres Kevlar-49 (0.365 GPa) and PBO (0.200 GPa) [163]. Compressive strength seems to increase with tensile strength and modulus, reflecting most likely differences in alignment for the samples tested so far. But it is necessary to determine clearer relations between structure and axial compressive properties; for related PAN-based CF, for example, axial compressive strength drops rapidly and significantly as tensile modulus is increased through improved graphitization and alignment, for MPP-based CF this change is less pronounced [110].

*3.8 Hierarchical structure of CNT fibre composites and unusual aspects of their mechanics*

The classical high-performance fibres have an all-solid "monolithic" structure with negligible porosity and BET surface area of less than 1 $m^2/g$ [76]. Polymer resin could not penetrate such fibres, and therefore, the fibre/matrix interface in reinforced matrices originates only at the external surface of a fibre and correlates with its outer diameter. In contrast, CNT fibres exhibit high porosity and well-developed surface area up to 250 $m^2/g$ depending on the synthesis conditions, which is thousands of times larger than that of conventional carbon fibres. More importantly, the inter-bundle gaps and pores of CNT fibres are accessible to various polymer molecules. When the CNT fibre are placed in contact with a liquid polymer matrix, polymer infiltration occurs instantly by capillary forces and can continue without applying pressure until the whole internal structure of a fibre is completely filled. An example of the spreading and wicking of a drop of viscous epoxy is shown in Figure 23 a. In the final CNTF composite, the polymer matrix surrounds the network of reinforcing elements (CNT bundles), and each constituent CNT fibre becomes a nanocomposite itself (Figure 23 b). In such hierarchical composite systems, the fibre-matrix interface is enormously large and can approach the total accessible surface area of the mesoporous CNT fibre.

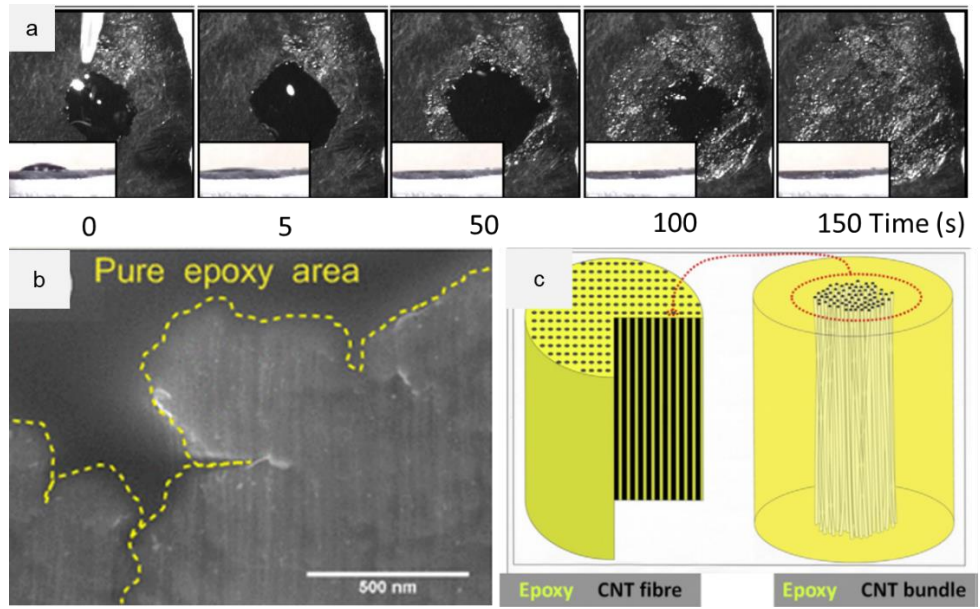

**Figure 23**. a) Optical micrographs of epoxy spreading into CNT fabric (inset shows side view). (adapted with permission from [140]. Copyright 2019 AIP Publishing LLC.); b) FIB-SEM cross section of the composites reinforced with CNT fibres. The image shows approximately a quarter of a well-infiltrated fibre and a region of epoxy matrix in-between of neighboring fibres (adapted with permission from [116]. Copyright 2016 Springer Nature); c) a schematic illustration of the hierarchical structure of CNT fibre composites where each fibre becomes essentially a nanocomposite with polymer matrix filling the inter-bundle space (adapted with permission from [171]. Copyright 2012 Elsevier Ltd.).

Numerous works have shown that the tensile strength and elastic modulus of a single fibre increase as a result of polymer infiltration and are no longer the same as those of the pristine as-synthesized fibre [28,172-175]. This is, again, an unusual situation compared to traditional high-performance fibres, whose tensile properties at the filamentary level are well-defined irrespectively of the absence or presence of surrounding polymer matrix.

The evaluation of *specific* modulus of the infiltrated fibres provides a direct evidence of enhanced stress transfer, distinctive from the more common volume-normalized mechanical properties, which could reflect the filling up of voids inside the material. Raman spectroscopy measurements have similarly confirmed that the ingress of polymer results in a larger stress transfer to the nanotubes in the composite fibres [140]. The downshift of Raman peak position upon application of a bulk tensile strain is an indication of the local stress/strain in the constituent nanotubes. When stress is uniformly distributed through the fibre, the rate of peak downshift with bulk strain can thus be related to the bulk young's modulus. In an effort to shed more light into polymer-enhanced inter-bundle stress transfer, Mas *et al.* compared Raman downshift rates (i.e. local modulus) for CNT fibres infiltrated with polymer matrices with different moduli. The results show that the enhancement in local modulus is correlated with the matrix young's modulus (Figure 24), that is, that stiffer matrices lead to a larger stress between nanotubes.

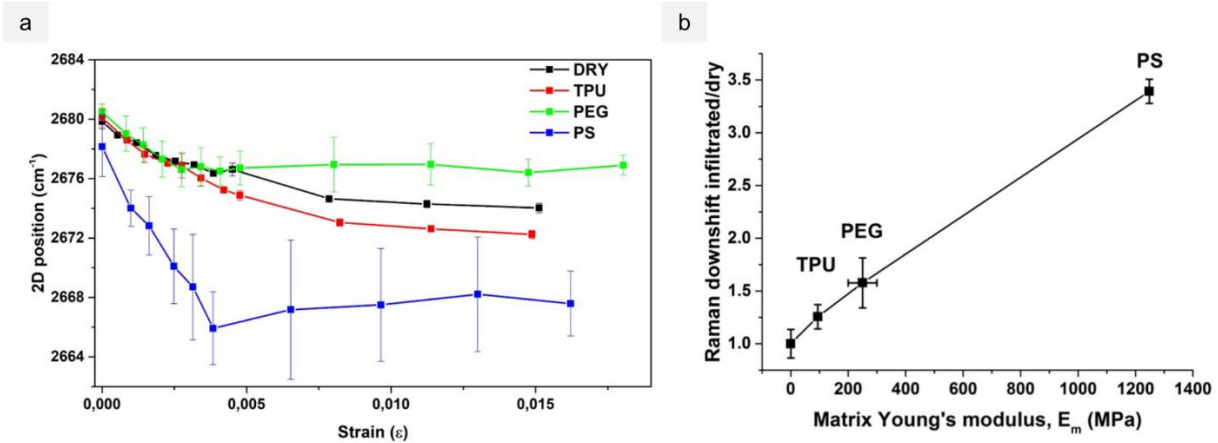

**Figure 24.** (a) 2D Raman peak position under tensile strain of dry individual CNT fibre (black), and embedded in thermoplastic polyurethane (TPU, red), polyethylene glycol (PEG, green) and polystyrene (PS, blue). (b) Plot of magnitude of 2D Raman peak position downshift per unit tensile strain, *versus*Young's modulus of the infiltrated polymer (images adapted with permission from [140]. Copyright 2019 AIP Publishing LLC.).

An interesting implication of the increase in CNT fibre modulus is that the composite longitudinal modulus cannot be *a priori* predicted using a rule of mixture solely with knowledge of the fibre and matrix moduli. But once the modulus of the fully infiltrated CNT fibre is known, this effective modulus could be used to determine composite properties. Figure 25 shows a schematic representation of the different structures and properties depending on the degree of infiltration and volume fraction, as well as supporting data. The challenge is then in being able to predict the effective modulus ($E_f$) of CNT fibres and fabrics for different polymer matrices. This will require being able to relate CNT fibre structure, properties of the fibre in the dry state, and matrix properties; a complex task.

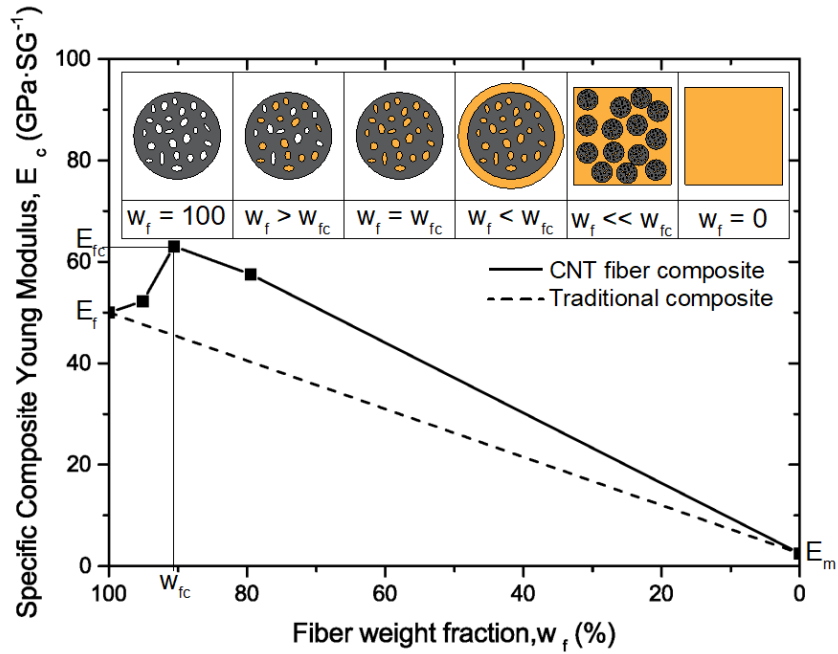

**Figure 25.** A schematic illustration of the specific tensile modulus of a CNT fibre-reinforced composite depending on the fibre weight fraction. $E_f$ at 100 wt.% corresponds to the modulus of the initial CNT fibre in the absence of any matrix, and $E_m$ is the modulus of pure polymer. $W_f$ is the weight fraction of fibres, $W_{fc}$ a critical weight fraction relative to the maximized modulus of the composite (reproduced with permission from [140] Copyright 2019 AIP Publishing LLC.).

The possibility to tune internal stress transfer in CNT fibres through polymer infusion represents and interesting opportunity to engineering properties of bulk composites. Coupled with the extraordinarily large interface with the matrix, this augurs interesting future avenues for improvement of toughness [176] and related properties in CNT fibre composites.

*3.9 Hybrid composites of CNT fibres and other macroscopic fibres*

A hybrid of CNT fibre with glass or carbon fibres is an attractive architecture to exploit the properties of CNT fibres in structural composites using relatively small fibre quantities. Thin (< 20 μm) CNT fabrics have a suitable format as interleaf layers for interlaminar reinforcement. Hybridization of conventional glass- and carbon fibre reinforced laminates with CNT fabrics thinner than an ultrathin prepreg have led to enhancements in interlaminar fracture toughness. Hybridization with thin unidirectional layer of CNT fibres (manufactured by winding of CNT fibres already condensed with ethanol) has also produced increases in longitudinal elongation-to-break by 9% without any negative effects on strength and stiffness of the laminates [146]. In recent studies in our group [177], we introduced low-alignment CNT fabrics in woven and unidirectional CF prepregs, respectively. The results show the critical importance of integrating the CNT fabric with the CF plies before densification to ensure improvements in interlaminar properties. When integrated in woven CF fabric, CNT fabric interleaves can lead to large increases of up to 64% in Mode-I fracture toughness, whereas in unidirectional prepregs they block CF bridging thus reducing fracture toughness in Mode-I fracture but increasing in Mode-II. This and other work are helping understand the key factors for adequate introduction of CNT fabric interleaves and the systems/configurations where they can lead to property improvements as well as those where they can be detrimental.

A hybrid architecture is very attractive as a vehicle for the introduction of CNT fibres and fabrics in structural composites using established industrial manufacturing methods. It will require overcoming engineering challenges such as: enabling and measuring complete polymer infusion, understanding curing kinetics and matrix chemistry inside the CNT fibres, and in general, producing large areas of uniform multifilament CNT fabrics.

**PART 4. Multifunctional composites based on CNT fibres**

Since the initial stages of the development of CNT fibre composites there has been a great interest in their potential to provide additional functions besides mechanical reinforcement. The longitudinal electrical [1] and thermal [118] conductivities of CNT fibres have been reported to be above that of metals, on a mass basis. Such properties are a result, in fact, of notable improvements in CNT fibre synthesis and assembly with the objective of increasing tensile modulus, again, through alignment of long highly conjugated few-layer nanotubes. In contrast with CF which present an inherent tradeoff between strength and transport properties, increases in CNTF strength and conductivity so far appear to go hand in hand. Indeed, improvements in composite mechanical properties consistently come with increases in electrical and thermal conductivity. From practical point of view, this unusual combination of properties unlocks applications such as structural elements with integrated heat management, power and data transmission, electromagnetic interference shielding and protection against lightning strike [178,179].

Other envisaged functions arise from the inherent piezo and chemoresistive properties of CNT fibres. Their piezoresistive behavior [180,181] is a consequence of their network structure and intrinsic molecular piezoresistance, whereas their chemoresistance is due to their low-dimensional electronic structure making them sensitive to the chemical environment [182]. These effects have been exploited in composites with integrated strain sensors [183], sensors that monitor polymer flow and curing during fabrication [184], and for structural health monitoring and damage detection [185,186], amongst others. Using the reinforcing fibres as embedded sensors avoids introducing invasive sensors, however, it remains to see if this benefit can be combined with sensing performance comparable to that of established technologies such as Bragg fibres.

Finally, another promising area of research is the combined use of CNT fibres as reinforcement elements and for energy management. Because of their unusual combination of high electrical conductivity, toughness, porosity and electrochemical stability, they are ideal current collectors for energy storing or harvesting devices with augmented mechanical properties. Replacing traditional metals with CNT fibre current collectors has been an important milestone in the development of flexible batteries, supercapacitors, and solar cells. Some examples are included in Figure 26. There is an active work on the progression from flexible to stretchable to structural devices, but the pace is often slow because of inherent limitations on the mechanical properties of active materials and electrolytes [187]. As an example of emerging developments, in Figure 26 we include an example of a structural supercapacitor composite made up of CNT fibres, CF, and a polymer electrolyte in epoxy vinyl ester.

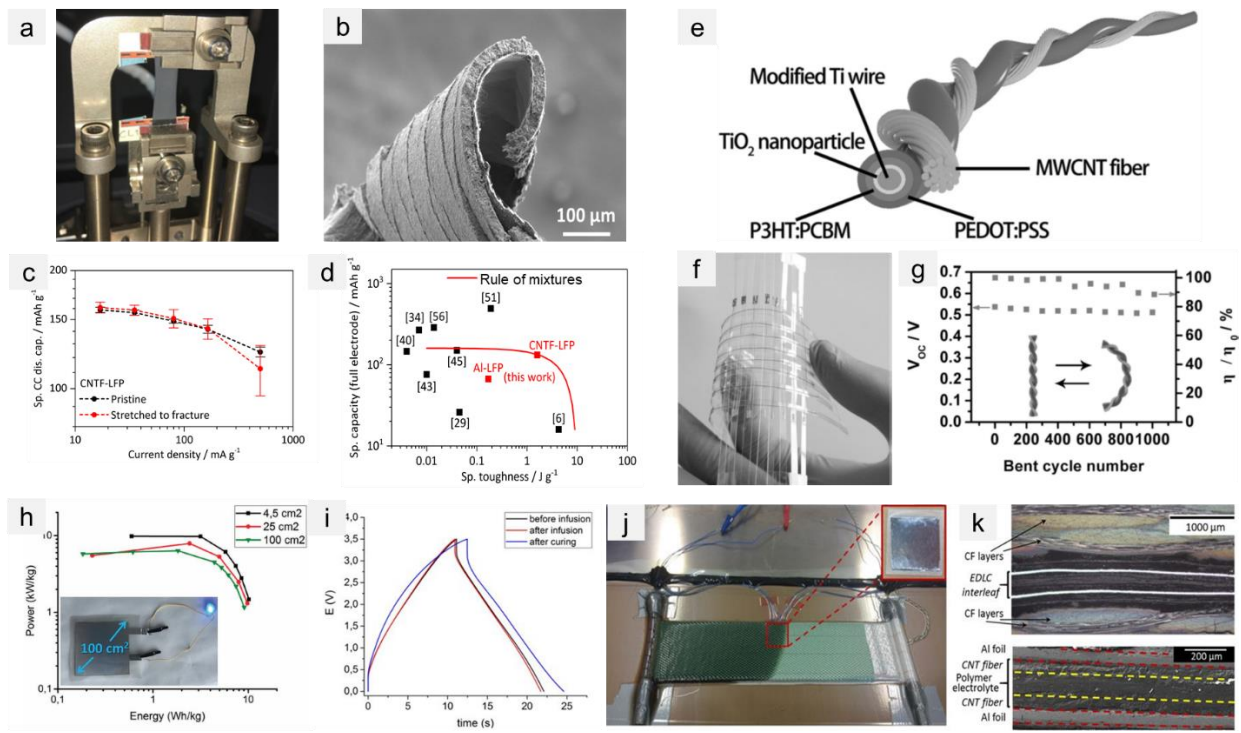

**Figure 26**. Examples of multifunctional composites: a) stretching of CNT fabric-based electrode, b) SEM image of the cross-section of a stretched electrode, c) discharge capacity of electrodes before and after stretching, d) specific capacity and toughness of developed electrodes (images a-d adapted with permission from [188]. Copyright 2019 WILEY-VCH Verlag GmbH & Co.); e) a schematic illustration of a wire-shaped polymer solar cell, f) an example of a PSC textile and g) the dependence of its energy conversion efficiency on bent cycle number (images e-g adapted with permission from [189] Copyright 2014 WILEY-VCH Verlag GmbH & Co.); and examples of structural electric-double layer supercapacitors (EDLC), including h) the Ragone plot with a photo of freestanding device of 100 $cm^2$ lighting a blue LED (inset) (images adapted with permission from [151]. Copyright 2017 WILEY-VCH Verlag GmbH & Co. ) and i) charge-discharge profiles of EDLC composite during fabrication by vacuum bag resin infusion, with the photographs of j) a CF/EDLC/CF laminate fabrication and k) its cross-section showing successful integration of EDLC in the laminate structure (images i-k adapted with permission from [190]. Copyright 2018 Springer Nature).

A particularly important filed of work is the development of batteries for the electrification of transport. Since their early years, CF were considered as anodes for lithium ion batteries [191]. More recently, there has been an interest in developing structural composite batteries [192], that are, structural composites that can simultaneously act as electrochemical energy storage elements and, thus, potentially lead to large weight reductions. The large surface area of CNT fibres makes them unsuitable as anodes in the pristine form, but is, on the other hand, beneficial for catalytic reactions in metal-air batteries. Overall, the fact that CNT fibres combine structural properties with a large surface for interfacial storage/transfer processes makes them ideal for emerging multifunctional energy storing composites.

CONCLUSION

Seen against a historical background, the development of CNT fibres is remarkably similar to that of CF, UHMWPE, and other high-performance fibres. Initially tensile properties were in the range of 0.15 GPa and 9-15 GPa strength and stiffness, but underwent a rapid increase to around 1.5 GPa and 70 GPa, respectively, after optimization of CNT synthesis and fibre spinning processes. LC-spun CNT fibres have strength of about 2.5 GPA (1.6 GPa/SG) and moduli of 250

GPa (160 GPa/SG) which is higher than that of aramids, Dyneema, and T300 and near the specific tensile modulus of Zylon HM and T1100. FCCVD-spun fibres have lower modulus but a fracture energy above that of CF and similar to polyaramids. These properties set CNT fibres firmly in the high performance range. Interestingly, in analogy to CF, the CNT fibres resulting from the different spinning processes (LC and FCCVD) seem to ultimately represent different grades from the point of view of resulting tensile properties, structure, and limiting factors.

For LC-spinning the dominant parameters are related to the starting CNT material and its ability to form nematic lyotropic LC phases, which restricts the choice of nanotubes in terms of aspect ratio and purity. Recent methods developed to screen fibre properties resulting from nanotubes of different morphology and from different suppliers, often available in very small quantities or at high cost, will be particularly useful for improvement in tensile strength and axial electrical conductivity. Together with the successful demonstration of wet-spinning from aqueous CNT dispersion, there will be increasing studies on structural composites from LC-spun CNT fibres.

Recent models show that tensile properties of FCCVD-spun fibres are strongly dominated by CNT alignment, which remains low. Post-spin drawing is therefore an important area of development. A better understanding of the state of FCCVD-spun CNT fibres during wet-drawing, to some extent similar to a gel fibre state, will be key for future developments.

Common to both types of fibres is the interest in cross-linking and other functionalization methods to tailor tensile properties. Here again, it will be critical to move from specific laboratory recipes to a better understanding of the relationship between surface chemistry, nature of the cross-linker, and the corresponding deformation mechanism in order to favour strength, modulus, and/or fracture energy.

An interest feature of CNT fibre materials, particularly those dry-spun, is the possibility to make fully integrated unidirectional multifilament fabrics directly by winding CNT aerogel filaments, without having to produce individual, densified fibres. This aspect of the materials is attractive from a technological point of view and opens interesting questions and challenges with respect to the integration of aerogel filaments to ensure cooperative load-bearing. It remains to see how easily the strategies and micromechanical models applied to single fibres can be applied to CNT fabrics.

From the point of view of composite fabrication and testing, the prevalent hurdle is the availability of large quantities of CNT fibres to manufacture composites using standard fabrication processes and then carry out standard testing protocols. For reference, a typical laminated structural composite has 8 layers of CF fabric each with an areal density of over 300 g/m$^2$. The typical areal density of CNTF for composites is typically closer to 30g/m$^2$. Some encouraging examples suggest the CNTF-composites can be fabricated using hot pressing or infusion methods. However, in the presence of CNT fabrics it is common practice to extend the infusion step in order to ensure full impregnation. Wicking studies on CNT fibres and fabrics show that capillary forces are effective to drive polymer infiltration and give an equivalent capillary radius of tens of nanometers, but the presence of a large mesoporosity is ultimately expected to slow down infusion. Further infusion studies are required to determine the competing effects of pressure gradient-driven and capillary-driven flow, and critically, to detect and characterize voids at the nanoscale and/or determine their effect on mechanical properties.

An important development will be the fabrication of CNT fibre composite lamina that enables determining extended mechanical properties, including in the transverse direction and under compression. This will enable prediction and simulation of properties of hybrid composites combining traditional fibres with CNT fibre lamina, which in view of the high fracture energy of CNT fibres could be interesting, for example, for impact protection.

An interesting test-ground for CNT fibres and fabrics is their introduction as interleaves between CF or GF lamina. These composites expose fabrication, characterization, and testing challenges similar to those expected for full CNT fibre composites. They also evidence how much is still unknown in terms of the structure and mechanics of CNT fibre/polymer composites.

Since their early developments, CNT fibres have been recognized as having an unusual combination of properties, making them ideal for multifunctional composites combining load-bearing capabilities with functions such as charge storage and conversion or sensing. With the emergence of integrated electronics and structural power, the opportunities for widespread implementation of CNT fibre-reinforced composites could, in fact, be motivated by applications that are not exclusively mechanical and, therefore, a broad scope must be kept.

The last 18 years since CNT fibres were first reported have seen continuous improvement in our understanding of this new material and overcoming the barriers for CNT fibre composite implementation. A large part of the development focus is still on the fibre level. Strong/stiff fibres seem difficult to produce, whereas "commercial" samples available as fabrics for composites have modest properties. For a traditional fibre this would indicate that composite development is still at a very early stage. However, there is a preliminary indication that CNT fabrics can be treated similarly to fibres in terms of their micromechanical properties and interaction with polymers. This is a paradigmatic change that implies that CNT fabric composites could be around the corner, provided current momentum in the field is kept and this fibre wisdom is applied to fabrics; possibly requiring a more coordinated effort between research laboratories, fibre spinning companies, composite manufactures, and end users.

AUTHOR CONTRIBUTIONS

The manuscript was written through contributions of all authors. All authors have given approval to the final version of the manuscript.

FUNDING SOURCES

A.M. acknowledges funding from the European Union's Horizon 2020 research and innovation programme under the Marie Skłodowska-Curie grant agreement 797176 (ENERYARN). J.J.V. is grateful for generous financial support provided by the European Union Seventh Framework Program under grant agreements 678565 (ERC-STEM) and Clean Sky-II 738085 (SORCERER JTI-CS2-2016-CFP03-LPA-02-11), and by the MINECO (RyC-2014-15115). The authors declare no competing financial interest.

ACKNOWLEDGMENT

The authors are grateful to Professor Alan Windle for extensive discussions of CNT fibres, especially when they had the privilege of being members of his research group.